\documentclass[a4paper,11pt]{article}    

\usepackage[latin1]{inputenc} 
\usepackage[T1]{fontenc}      
\usepackage{authblk}
\usepackage{graphicx}

\usepackage{amsmath}
\usepackage{amssymb}
\usepackage{verbatim}
\usepackage{appendix}

\usepackage{chicago}
\usepackage{color}
\newcommand{\red}[1]{#1}

\newcommand{\pa}[1]{\left(#1\right)}      
\newcommand{\pasq}[1]{\left[#1\right]}      
\newcommand{\pacu}[1]{\left\{#1\right\}}      
\newcommand{\Ex}[1]{\left\langle#1\right\rangle}

\newcommand{\dd}{\mathrm{d}} 

\newcommand{\bx}{\mathbf{x}}
\newcommand{\bk}{\mathbf{k}}
\newcommand{\bi}{\mathbf{i}}
\newcommand{\bj}{\mathbf{j}}
\newcommand{\bt}{\mathbf{t}}
\newcommand{\bu}{\mathbf{u}}

\newcommand{\balpha}{\boldsymbol{\alpha}}
\newcommand{\bgamma}{\boldsymbol{\gamma}}
\newcommand{\tepsilon}{\widetilde{\epsilon}}
\newcommand{\btepsilon}{\widetilde{\boldsymbol{\epsilon}}}
\newcommand{\bzeta}{\boldsymbol{\zeta}}

\renewcommand{\Re}{\mathbb{R}}
\newcommand{\N}{\mathbb{N}}

\def\deltabar{{\mathchar '26\mkern -10mu\delta}} 

\newtheorem{theorem}{Theorem}
\newtheorem{definition}{Definition}
\newtheorem{notation}{Notation}
\newtheorem{proposition}{Proposition}
\newtheorem{corollary}{Corollary}
\newtheorem{example}{Example}
\newtheorem{property}{Property}
\newtheorem{remark}{Remark}

\newenvironment{myproof}[2] {\paragraph{{\bf Proof of {#1}~{#2}:}}}{\hfill$\square$}

\title{Propagation of spiking moments in linear Hawkes networks}

\author[1]{Matthieu Gilson}
\author[2,3]{Jean-Pascal Pfister}

\affil[1]{Universitat Pompeu Fabra, Barcelona, Spain}
\affil[2]{Institute of Neuroinformatics and Neuroscience Center Zurich, University of Zurich/ETH Zurich, Zurich, Switzerland}
\affil[3]{Department of Physiology, University of Bern, Bern, Switzerland}

\begin{document}

\maketitle

\begin{abstract}
The present paper provides exact mathematical expressions for the high-order moments of spiking activity in a recurrently-connected network of linear Hawkes processes. It extends previous studies that have explored the case of a (linear) Hawkes network driven by deterministic intensity functions to the case of a stimulation by external inputs (rate functions or spike trains) with arbitrary correlation structure. Our approach describes the spatio-temporal filtering induced by the afferent and recurrent connectivities (with arbitrary synaptic response kernels) using operators acting on the input moments. This algebraic viewpoint provides intuition about how the network ingredients shape the input-output mapping for moments, as well as cumulants. We also show using numerical simulation that our results hold for neurons with refractoriness implemented by self-inhibition, provided the corresponding negative feedback for each neuron only mildly alters its mean firing probability.
\end{abstract}


\section{Introduction}

Immense efforts in neuroscience have been invested in measuring neuronal activity as well as the detailed connectivity between neurons. Such studies have been too often conducted separately, despite the fact that neuronal activity and synaptic connectivity are deeply intertwined. Indeed, the synaptic connectome determines the neuronal activity, while the latter reshapes the connectome through activity-dependent plasticity. To better understand the intricate link between activity and connectivity at the neuronal level, it is important to build tractable network models that relate one to the other. 

\begin{figure}
        \includegraphics[scale=1]{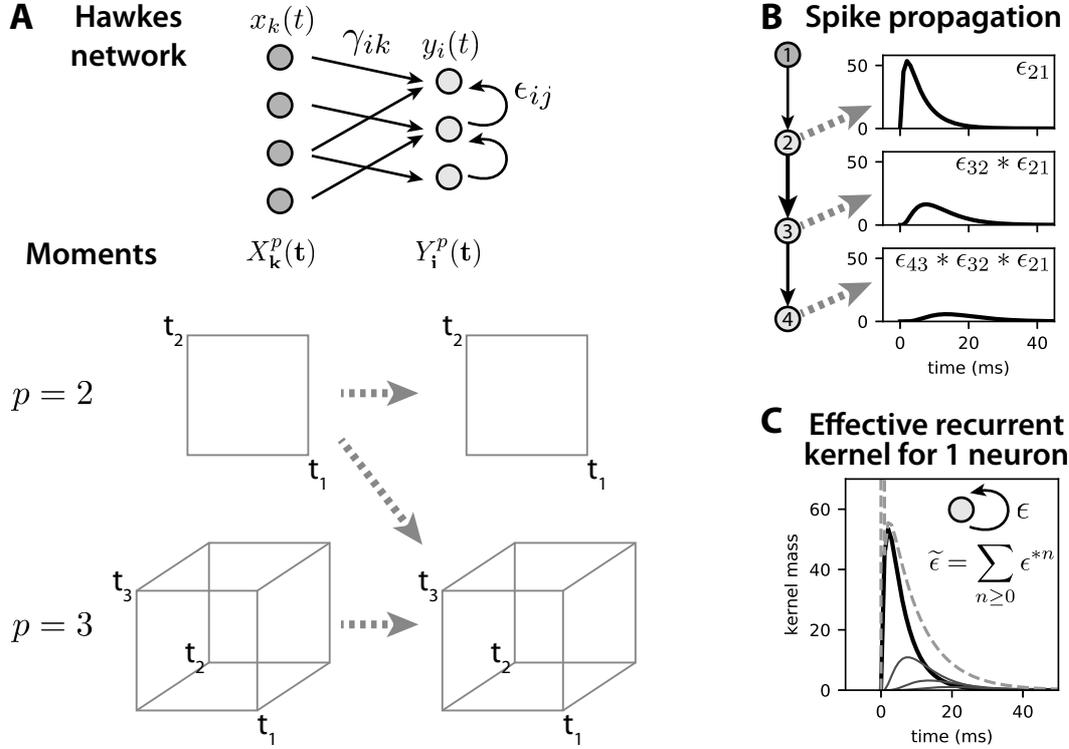}
        \caption{ \label{fig1_schem}
        {\bf Overview of the present study.}
        {\bf A:} This schematic diagram at the top represents a Hawkes network, where nodes are the individual neurons that emit (or fire) spikes, borrowing the terminology in neuroscience. The afferent and recurrent connectivity are described by the kernel functions $\gamma_{ik}$ and $\epsilon_{ij}$, respectively. 
        The goal of the present work is the characterization of the mapping between the moments of the input and output spike trains (i.e.\ their correlation structure). They are represented by the matrices and cubes, respectively representing the second- and third-order moments that are formally tensors with ``spatial'' coordinates (over neurons) and temporal variables. The dashed gray arrows represent cross-order contributions from the input to the output moments.
        {\bf B:}
        This diagram depicts the average firing intensity of the downstream neurons in the output population due to a spike fired by the dark gray neuron (assuming that $\epsilon_{ij} = 0$ for $i\neq j+1$). The black curves represent the increase in average conditional intensity $\Ex{\nu_i}_{y}$ at the light gray neurons following the spike in neuron 1, which is given by the convolution of the synaptic kernels $\epsilon_{ij}$ of the corresponding connections.
        {\bf C:}
        Similar diagram to panel B for a neuron with a self-connection with kernel $\epsilon$ (thick solid black curve). The effective recurrent kernel $\tepsilon$ (dashed gray curve) is given by the superposition of $\epsilon$ with its self-convolutions (thin solid gray curves). 
        Generalizing, we can calculate the effective recurrent kernel in the multivariate case for interconnected neurons. It corresponds to the Green function of the network in the context of linear dynamics.
        }
\end{figure}

This paper examines how the neuronal activity is determined by the synaptic connectivity in a network. 
More precisely, we investigate how the spiking statistics ---described via statistical moments or cumulants--- propagates from an input population of neurons to an output population of recurrently-connected neurons, see Fig.~\ref{fig1_schem}A.
Their firing probability depends on upstream neurons, as represented in Fig.~\ref{fig1_schem}B. \red{To formalize this relationship, one needs to decide on a model for the neuronal dynamics.}

From the large class of existing neuronal models, we chose the simplest possible model in order to remain tractable. Indeed, detailed biophysical models such as the Hodgkin-Huxley model or the conductance-based models nicely describe the membrane potential dynamics around the action potential but are harder to study when embedded in a network.
Their complexity often requires numerical simulation for their study and optimization strategies are still under debate, see for example~\cite{Brette_JCN_2007,Lai_PRE_2017}.
It can be argued that only the timing of the action potential that matters for the postsynaptic neuron, which motivates our choice for point processes where action potentials are events.
A somewhat simple class of neuronal model is the so-called spike-response model~\cite{Gerstner_book_2002}, also known as exponential Poisson model or generalized linear model, GLM~\cite{Pillow_Nature_2008}. It is worth noting that such simple models often provide the best fit to data in terms of predicting the timing of action potentials for a single neuron driven by a controlled input current~\cite{Gerstner_Science_2009}. \red{Formally, such spiking neuron models correspond to non-linear versions of Hawkes processes when coupled together in a network.
For tractability purpose, we model here the spiking activity using a linear Hawkes process~\cite{Hawkes_JRSB_1971,Hawkes_Biom_1971}, also known as (linear) Poisson neurons~\cite{Kempter_PRE_1999}.
In the following we refer to the multivariate linear Hawkes process as Hawkes network.
}

Despite the obvious limitations of the linearity assumption (e.g.\ it precludes strong refractoriness or inhibition), Hawkes' formalism has been extensively used to model recurrent spiking network~\cite{Gilson_FCN_2010,Pfister_FCN_2010,Mei_NIPS_2017}. Indeed the reason for its wide adoption is its analytical tractability which precisely comes from the linear assumption. Beside neuroscience, the Hawkes process has been used in several other disciplines such as  artificial intelligence~\cite{Etesami_UAI_2016}, seismology~\cite{Le_IEEE_2018,Lima_arxiv}, epidemiology~\cite{Saichev_EPJB_2013} and finance~\cite{Errais_SIAM_2010,Bacry_MML_2015}. Due to the event-like nature of its activity, intrinsic correlations arise and reverberate as echoes induced by the recurrent connectivity. Here we build upon Hawkes' results that describe the propagation of second-order correlations for mutually exciting point processes~\cite{Hawkes_JRSB_1971,Hawkes_Biom_1971} and extend them to higher orders. 

The vast majority of studies focuses on the first and second orders of spiking statistics~\cite{Hawkes_JRSB_1971,Hawkes_Biom_1971,Gilson_FCN_2010,Bremaud_AAP_2005,Tannenbaum_PRE_2017}. Up to our knowledge, only two recent studies have investigated higher-order cumulants~\cite{Jovanovic_PRE_2015,Ocker_PCB_2017}. In the earliest~\cite{Jovanovic_PRE_2015}, the authors derived a recursive algorithm based on the theory of branching Hawkes processes to calculate the cumulants for the spiking activity. 
The second study~\cite{Ocker_PCB_2017} relies on path-integral representation to explore the cumulants, which are closely related to moments, for Hawkes process with possible non-linearities. If the path-integral representation derived from field theory is adequate to tackle non-linearities, it requires approximations with a cumulant closure to obtain self-consistency equations.
A common limitation to both formalisms is that they provide little intuition about how the moments may propagate in neuronal networks, which we aim to address here focusing on its geometrical aspect ---as will become clearer later, see also the representation of moments in Fig.~\ref{fig1_schem}A. Importantly, the case of neurons stimulated by inputs with an arbitrary correlation structure has not been explored yet for larger-than-second orders, which is a focus of the present study.

A first motivation is that, although pairwise correlations have been argued to be sufficient to represent experimental data~\cite{Barreiro_FCN_2014}, this view has been recently challenged and mechanisms related to higher-order correlations have been found to improve descriptive statistical models~\cite{Shimazaki_SciRep_2015}.
In dynamic neuron models, even though population mean-field dynamics can be captured by non-spiking models~\cite{Helias_NJP_2013,Grytskyy_FCN_2013}, networks with realistic sizes exhibit finite-size effects in their pairwise correlations~\cite{Albada_PCB_2015}. 
Moreover, there is accumulating evidence in biology that spike trains convey information in their correlated activity~\cite{Dettner_NatC_2016}.
This calls for analytical techniques to evaluate the interplay between spiking correlated activity at arbitrary orders (as measured by moments or cumulants) and network connectivity, as was done recently for binary neurons~\cite{Dahmen_PRX_2016}.

A motivation for investigating higher-than-second orders of correlations in Hawkes networks comes from the study of spike-timing dependent plasticity (STDP). The established formula~\cite{Hawkes_JRSB_1971} is sufficient to analyze in recurrently-connected networks the effect of the so-called pairwise STDP: As the synaptic weights between neurons are modified depending on the time difference between input and output spikes, the overall effect can be captured by the spiking covariances~\cite{Gilson_BC_2009a,Gilson_BC_2009b,Pfister_FCN_2010}. However, the more elaborate model of triplet STDP~\cite{Pfister_JNS_2006,Gjorgjieva_PNAS_2011} requires the knowledge about the third order of the spike statistics, involving input-output-output spikes. To gain intuition, a key is understanding how the synaptic connectivity shapes the input correlation structure in a network as illustrated in Fig.~\ref{fig1_schem}A.

This led us to investigate a general solution for the spatio-temporal correlation structure via moments of arbitrary orders in Hawkes processes as a function of the moments in the input population. Our results are structured around three theorems. The first one describes how moments (of arbitrary orders) propagate in feedforward networks, thereby generalizing the results by~\cite{Kempter_PRE_1999}. The second theorem describes the effect of recurrent connectivity within the output population, extending~\cite{Gilson_BC_2009b,Pfister_FCN_2010}. 
Our calculations assume that the firing intensities of the neurons remain positive at all times. We discuss the limitations of this assumption in an example with self-inhibition.
The last theorem translates the mappings for moments into mappings for cumulants, in line with recent work~\cite{Jovanovic_PRE_2015,Ocker_PCB_2017}.

\section{Results}

Let us consider an input population of $m$ neurons whose spiking activity is denoted by the vector of functions%
\footnote{Depending on the context, $x$ can be a given (deterministic) spike train or a random variable. For notational convenience, we decided to use the same symbol $x$ for both cases. The same holds true for $\lambda$. For example, in the definition of the Hawkes process, $x$ and $\lambda$ are given (deterministic) whereas later on, we compute the statistics of the output population by averaging over the random variables $x$ and $\lambda$ such as in Eq.~\ref{eq:mom_out}.} %
$x(t) = (x_1(t), \cdots, x_m(t))$ where $x_j(t)$ is a superposition of Dirac deltas at spike times, i.e.\ $x_j(t) = \sum_{f}\delta(t-t_{j,f}^x)$ and $t_{j,f}^x$ is the $f^{\rm th}$ firing time of the input neuron $j$. As illustrated in Fig.~\ref{fig1_schem}A, this input population together with some driving intensity function $\lambda(t) = (\lambda_1(t),\cdots,\lambda_n(t))$ feed a network (output) population of $n$ neurons whose activity is denoted by $y(t) = (y_1(t), \cdots, y_n(t))$, which are also a superposition of Dirac deltas, i.e.\ $y_i(t) = \sum_f\delta(t-t_{i,f}^y)$. A Hawkes process formalizes how the output spikes $y(t)$ are generated from the history of input spikes $\mathcal{F}^x_t = \{x(s)|s<t\}$, the history of output spikes $\mathcal{F}^y_t = \{y(s)|s<t\}$ \footnote{Formally, the filtrations $\mathcal{F}^{x}_t$ and $\mathcal{F}^{y}_t$ are $\sigma$ fields.} and the driving intensity $\lambda(t)$. 
In the literature Hawkes processes are often defined by a $n-$dimensional counting process $N^y(t) = (N^y_1(t),\dots,N^y_n(t))$, where $N^y_i(t)$ gives the number of spikes from $0$ to $t$ for the network neuron $i$, i.e.\ $N^y_i(t)= \int_0^t y_i(t') \, \dd t'$ or equivalently $y_i(t) = \dd N^y_i(t)/\dd t$.
\red{The heart of Hawkes' theory lies in the conditional intensity $\nu_i(t)$ that determines the probability of an event (here a spike), via the increment of the counting process $\dd N^y_i(t) = N^y_i(t + \dd t) - N^y_i(t)$ for neuron $i$ in an infinitesimally small bin size $\dd t$:
\begin{equation} \label{eq:cond_int}
    \nu_i(t) \ \dd t = \Pr\Big(\dd N^y_i(t) = 1 | \mathcal{F}^y_t , \mathcal{F}^x_t, \lambda(t) \Big) \ .
\end{equation}
Alternatively, the increment $\dd N_i^y(t)$ at each time can be seen as resulting from a Poisson process with the conditional intensity $\nu_i(t)$:
\begin{equation}
    \dd N^y_i(t) \sim \mathrm{Poisson}\big( \nu_i(t) \dd t \big) \ .
\end{equation}
For further detail, see~\cite{Bremaud_AP_1996} or~\cite[Section~6.3]{Daley_1988} for the general theory on related point processes constructed using conditional intensities.
A consequence of this property is that $\Ex{\dd N^y_i(t)} = \nu_i(t)dt$ at all times, hence $\Ex{y_i(t)} = \Ex{\dd N^y_i(t)/\dd t} = \nu_i(t)$. Here the conditional expectation denoted by the angular brackets is taken on the increment $\dd N^y_i(t)$ ---or equivalently on $y_i(t)$--- at time $t$ given the history of $x$ and $y$ and the value of $\lambda$ at time $t$ (as in Eq.~\eqref{eq:cond_int}). In the following, we denote the conditioning variables as subscripts of the angular brackets, for instance $\Ex{y_i(t)}_{y,x,\lambda}$. The interested reader can find further detail referred to~\cite[Section~7.2, Example~7.2]{Daley_1988}.
}

In this paper we use the following definition for Hawkes process:
\begin{definition}[Hawkes Process] \label{def:hawkes}
The Hawkes process is a $n$-dimensional point process $y(t)$ whose conditional intensity $\nu: \Re \rightarrow \Re_+^n$ is driven by a time-dependent intensity $\lambda: \Re \rightarrow \Re_+^n $ and depends upon both the past input spiking activity $\mathcal{F}^x_t$ and its own past spiking activity $\mathcal{F}^y_t$ :
\begin{equation} \label{eq:Hawkes}
    \nu_i(t) = \lambda_i(t)  + \pa{ \gamma_{ik} \ast x_k }(t) + \pa{ \epsilon_{ij} \ast y_j }(t) \ ,
\end{equation}
where $\gamma = \{\gamma_{ik}\}_{i,k=1}^{n,m}: \Re \rightarrow \Re^{n\times m}_+$ is a matrix of ``synaptic'' kernels $\gamma_{ik}: \Re \rightarrow \Re_+$ that describe the causal effect from the input neuron $x_k$ on the network neuron $y_i$.  These functions are equal to zero for all $t \leq 0$. Similarly $\epsilon = \{\epsilon_{ij}\}_{i,j=1}^{n,n}: \Re \rightarrow \Re^{n\times n}_+$ is a matrix of kernels $\epsilon_{ij}: \Re \rightarrow \Re_+$, each corresponding to the recurrent interraction from neuron $y_j$ to neuron $y_i$. 
\end{definition}

The definition in Eq.~\eqref{eq:Hawkes} has the implicit assumption that the right-hand side is always non-negative, which is satisfied for our choice of $\gamma$ and $\epsilon$ here. Recall that this is not guaranteed in general when these kernels have negative values. In such cases one can apply a rectifying function to ensure that $\nu_i(t) \geq 0$, which brings nonlinearity in the formalism~\cite{Bremaud_AP_1996,Galves_JSFS_2016,Ocker_PCB_2017,Gao_SPA_2018,Raad_arXiv_2018}.

Note that the convolution operator $\ast$ is a matrix convolution (see Eq.~\eqref{eq:def_conv} below). Note also that in this paper we omit the summation symbol in line with Einstein's convention for tensor calculus. 

\begin{remark}[Atomic contributions and contraction of indices] \label{rem:contraction}
Note that for an infinitesimally small $\dd t$, the increment $\dd N_i^y(t)$ for the output neuron $i$ and $\dd N_k^x(t)$ for the input neuron $k$ can take only 2 values: 0 or 1. In that case, we have for any $p \in \N_+$
\red{
\begin{eqnarray} \label{eq:dNp}
    \pa{\dd N_i^{y}(t)}^p & = & \dd N_i^{y}(t)
    \ , \\
    \pa{\dd N_k^{x}(t)}^p & = & \dd N_k^{x}(t) 
    \nonumber
\end{eqnarray}
}
Following, atomic contributions arise from the point-process nature of spike trains when taking expectations of products of input spike trains $x_i(t)$ or output spike trains $y_i(t)$  for all possible redundancies in the time variables together with the ``spatial'' coordinates.
\end{remark}

Remark~\ref{rem:contraction} can be directly used to compute moments for independent Poisson neurons. For example, the second order moment for the an input population of independent Poisson neurons yields
\begin{eqnarray} \label{eq:ex_contraction}
  \Ex{ x_{k_1}(t_1) x_{k_2}(t_2) }_x
  &=& 
  \left\{
  \begin{array}{l l}
   \Ex{\frac{\dd N^x_{k_1}(t_1)}{\dd t}\frac{\dd N^x_{k_2}(t_2)}{\dd t}}_x
  &
  \mathrm{if~} k_1 \neq k_2 \mathrm{~or~} t_1 \neq t_2
  \\ 
  \Ex{\pa{\frac{\dd N^x_{k_1}(t_1)}{\dd t}}^2}_x
  &
  \mathrm{if~} k_1 = k_2 \mathrm{~and~} t_1 = t_2
  \end{array}
  \right.
  \nonumber\\
  &=&
  \Ex{ x_{k_1}(t_1) }_x \Ex{ x_{k_2}(t_2) }_x
  +
  \Ex{x_{k_1}(t_1)}_x\delta_{k_1k_2}\delta(t_2-t_1)
  \ ,
\end{eqnarray}
where we use Eq.~(\ref{eq:dNp}) for the second term corresponding to $t_1=t_2$ and $k_1=k_2$. Note that the second term of Eq.~(\ref{eq:ex_contraction}) dominates the first term, which is related to the Dirac delta function $\delta(t_2-t_1)$ due to the limit $\dd t \rightarrow 0$ in the denominator that remains after simplifying $[ \dd N^x_{k_1}(t_1) ]^2 = \dd N^x_{k_1}(t_1)$.
In the remainder we refer to terms involving Kronecker deltas $\delta_{k_1 k_2}$ and Dirac deltas $\delta(t_2-t_1)$ as contractions of indices (here for $1$ and $2$).

We extend the standard convolution to a matrix form, which involves a matrix multiplication as in Eq.~\eqref{eq:Hawkes}. 
\begin{definition}[Matrix convolution] \label{def:mat_convol}
For the kernel matrix $\epsilon$ and vector $y$, the $i^\mathrm{th}$ element of the matrix convolution is given by 
\begin{equation} \label{eq:def_conv}
    \pa{ \epsilon_{ij} \ast y_j }(t)
    =
    \sum_{j=1}^{n} \int_0^\infty \epsilon_{ij}(u) y_j(t-u) \dd u
    \ .
\end{equation}
\end{definition}
%

\begin{notation}[Moments of order $p$]
Let $\bk = (k_1,\cdots,k_p)$ denote a set of $p$ coordinates $k_r \in I_m = \{1, \cdots, m\}$. The moment of order $p$ of the input population evaluated at times $\bt = (t_1,\cdots,t_p)$ is defined as
\begin{equation} \label{eq:mom_in}
    X^p_{\bk}(\bt)
    =
    \Ex{ \prod_{r=1}^p x_{k_r}(t_r) }_{x}
    \ .
\end{equation}
Similarly, the moment of order $p$ of the output population for the coordinates $\bi = (i_1,\cdots,i_p)\in I_n^p$ and the time variables $\bt = (t_1,\cdots,t_p)$ is defined as
\begin{equation} \label{eq:mom_out}
    Y^p_{\bi}(\bt)
    =
    \Ex{\prod_{r=1}^p y_{i_r}(t_r)}_{y,x,\lambda}
    \ .
\end{equation}
Note that the mathematical expectation corresponds to three sources of stochasticity, as indicated by the superscript. Note that, due to the recurrent connectivity, the dependency of $y$ on itself also concerns the past activity.
\end{notation}

\begin{remark}[Symmetry of moments] \label{rem:sym_mom}
The moments $X^p_{\bk}(\bt)$ and $Y^p_{\bi}(\bt)$ have many symmetries. For the example of the input moments, any permutation $\Pi$ of $I_p$ such that the transformed coordinates $\Pi(\bk) = (k_{\Pi(1)}, \cdots, k_{\Pi(p)})$ and $\Pi(\bt) = (t_{\Pi(1)}, \cdots, t_{\Pi(p)})$ leaves $X^p_{\bk}(\bt)$ invariant: 
\begin{equation} \label{eq:sym_mom}
    X^p_{\Pi(\bk)}\big(\Pi(\bt)\big) = X^p_{\bk}(\bt)
    \ .
\end{equation}
\end{remark}
%

\begin{definition}[Generalized spatio-temporal delta function] \label{def:gen_delta}
Let $\deltabar_{\bk}(\bt)$ be the generalized delta function defined for the set of coordinates $\bk = (k_1,\cdots,k_p)$ and times $\bt = (t_1,\cdots,t_p)$, which combines the Kronecker and Dirac delta functions as
\begin{equation} \label{eq:def_grl_delta}
    \deltabar_{\bk}(\bt)
    =
    \left\{ \begin{array}{ll}
        1 & \mathrm{~if~} p \in \{0, 1\} \ ,
        \\
        \prod_{r=2}^p \delta\pa{t_{r-1}-t_r} & \mathrm{~if~} k_1 = \cdots = k_p \mathrm{~and~}  p \geq 2 \ ,
        \\
        0 & \mathrm{~otherwise.}
    \end{array} \right.
\end{equation}
Note that for $p=2$, one recovers the product of the standard Kronecker delta with the Dirac delta: $\deltabar_{k_1,k_2}(t_1,t_2) = \delta_{k_1,k_2}\delta(t_1-t_2)$. Note also that when the lower index $\bk$ is omitted, we will assume that $k_1=\cdots = k_p$ (i.e.\ single neuron case). 
\end{definition}
%

\begin{example}[Moment for a single spike train with oscillatory intensity] \label{ex:ex_in}

Before presenting the general result, we provide an illustrative example to fix ideas and help the reader with concepts and notation.

\underline{Case $p = 2$:}

For a single (input) neuron driven by a deterministic intensity function $\mu$, the contraction in the 2nd-order moment corresponds to the condition $t_1 = t_2$ without ``spatial'' coordinates here, simplifying Eq.~\eqref{eq:ex_contraction}:
\begin{equation} \label{eq:dev_mom2}
    \Ex{ x(t_1) x(t_2) }_x
    =
    \mu(t_1) \mu(t_2) +  \mu(t_1) \, \delta(t_2-t_1)
    \ .
\end{equation}

\underline{Case $p = 3$:}

The 3rd-order moment for a single spike train is given by
\begin{eqnarray} \label{eq:dev_mom3}
    \Ex{ x(t_1) x(t_2) x(t_3) }_x
    & = & 
    \mu(t_1) \mu(t_2) \mu(t_3) + \mu(t_1) \delta(t_2-t_1) \mu(t_3) + \delta(t_1-t_3) \mu(t_1) \mu(t_2)
    \nonumber \\
    & & + \mu(t_1) \mu(t_2) \delta(t_3-t_2) + \mu(t_1) \delta(t_2-t_1) \delta(t_3-t_1) 
    \ .
\end{eqnarray}
This expression exhibits two ``extreme'' cases where all time variables are equal $t_1 = t_2 = t_3$ corresponding to the two Dirac delta $\delta(t_2-t_1) \delta(t_3-t_1) = \deltabar(t_1,t_2,t_3)$ for the partition $\big\{ \{1, 2, 3\} \big\}$, and where they are all distinct giving $\mu(t_1) \mu(t_2) \mu(t_3)$ for $\big\{ \{1\}, \{2\}, \{3\} \big\}$. In addition, the three remaining terms involve a contraction for 2 out of the 3 variables.

\underline{Numerical simulation:}

\begin{figure}
        \includegraphics[scale=0.94]{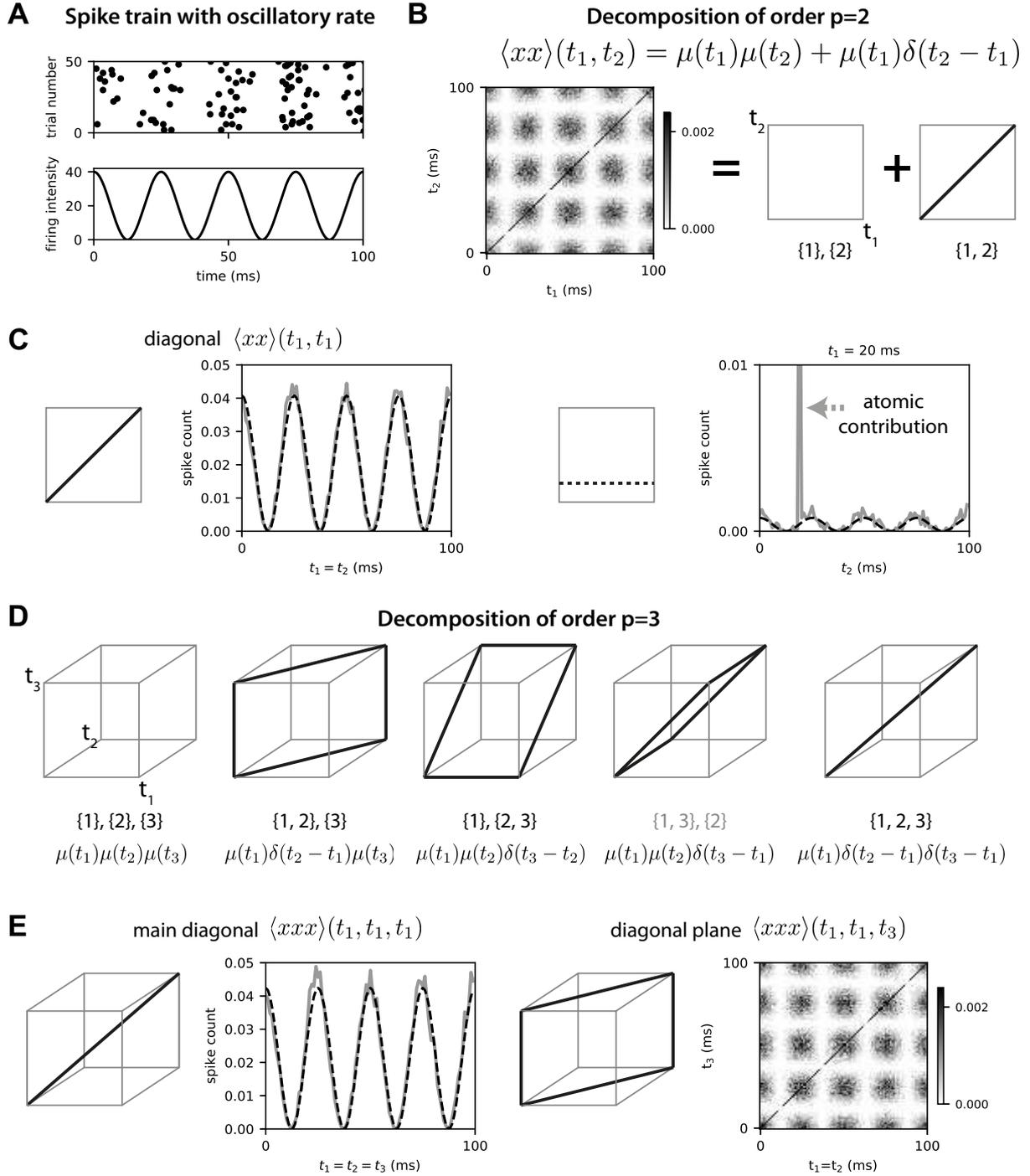}
        \caption{ \label{fig2_mom_cmplt}
        {\bf Input moments for a spike train.}
        {\bf A:} Spike raster (top plot) for 50 simulations using a driving oscillatory intensity (bottom plot).
        {\bf B:} 2nd-order moment (left plot) averaged over 10000 simulations, where darker pixels indicate a higher spike density. The middle and right diagrams illustrate the decomposition into a contribution due to rate correlation (co-fluctuations) and to atomic contributions (diagonal in thick black), respectively. Each contribution corresponds to a partition of $I_2 = \{1, 2\}$, as indicated below. 
        {\bf C:} Example slices of the moment in panel B as indicated by the solid/dotted lines in the diagrams, along the diagonal $t_1 = t_2$ (left diagram and plot) and for a fixed $t_1$ (right diagram and plot; here the atomic contribution is not represented for the theoretical prediction).
        The prediction curves (dashed) are calculated using Eq.~\eqref{eq:prop1_input}. Note the difference in scaling for the y-axis.
        {\bf D:} Decomposition of the 3rd-order moment using the partitions of $I_3 = \{1, 2, 3\}$, similar to panel B. The thick black lines indicates the diagonal planes and diagonal line for all possible contractions.
        {\bf E:} The left diagram and plot correspond to the main diagonal of the 3rd-order moment with $t_1 = t_2 = t_3$. The right diagram and plot correspond to the diagonal plane with $t_1 = t_2$.
        }
\end{figure}

Fig.~\ref{fig2_mom_cmplt} illustrates the moments for $p = 2$ and $3$ with a single spike train driven by a driving oscillatory intensity. Note that ``spatial'' coordinates $\bk$ in the above equation are simply ignored, together with the Kronecker deltas. Fig.~\ref{fig2_mom_cmplt}B, C and E highlight the atomic contributions along the various ``diagonals'' where the time variables coincide. Away from those subspaces, the spike densities are much lower, as can be seen in the scaling of values in the middle and right plots of Fig.~\ref{fig2_mom_cmplt}C and E.
Note that the main diagonal for $p = 3$ is slightly larger than that for $p = 2$, as autocorrelation effects cumulate.

\end{example}

\begin{proposition}[Moments for inputs driven by deterministic intensity functions] \label{prop:prop_input}
Let $\mathcal{P}_p = \mathcal{P}(I_p)$ denote the set of all partitions $\Phi$ of the set $I_p = \{1,\cdots,p\}$.
If the input neurons are independent from one another and driven by intensities $\mu_k(t)$, then the input moment of order $p$ with coordinates $\bk = (k_1,\cdots,k_p)$  at times $\bt = (t_1,\cdots,t_p)$ can be expressed as
\begin{equation} \label{eq:prop1_input}
    X^p_\bk(\bt)
    =
    \sum_{\Phi \in \mathcal{P}_p} \prod_{S \in \Phi} \deltabar_{\bk_S}(\bt_S) \ \mu_{k_{\check{S}}}(t_{\check{S}})
    \ ,
\end{equation}
where $S$ spans the disjoint subsets of $\Phi$ whose union is $I_p$, with $\bk_S = \{k_{r}, r \in S\}$ and $\bt_S = \{t_{r}, r \in S\}$.
In addition, each driving intensity $\mu$ appears with a representative index, here taken as the minimum $\check{S} = \min(S)$.
Recall the convention $\deltabar_{\bk_S}(\bt_S) = 1$ when $S$ is a singleton. 
\end{proposition}

\begin{remark}
The grouping of indices from a given subset $S$ in Eq.~\eqref{eq:prop1_input} is a direct consequence of the \emph{contraction} highlighted in remark~\ref{rem:contraction}, resulting in an \emph{atomic contribution} where the paired spatial coordinates and temporal variables related to $S$ are involved in the generalized delta function $\deltabar$. 
\end{remark}

\begin{myproof}{Proposition}{\ref{prop:prop_input}} \label{proof:proof1}

Eq.~\eqref{eq:prop1_input} can be obtained using the moment generating function via its $p^\mathrm{th}$ derivative for order $p$, as was done in previous work in similar contexts~\cite[Section~5.2]{Ocker_PCB_2017,Daley_1988}.
Here we provide a proof by induction, which highlights the key observation that every combination of contractions can be described by a partition.

Let assume that Eq.~\eqref{eq:prop1_input} is valid for all orders $2 \leq p' \leq p-1$. Now considering the order $p$ with given coordinates $\bk$ and time variables $\bt$ in $X_\bk(\bt)$, we denote by $S^*$ the set of order indices in $I_{p-1}$ such that coordinates and times are identical to their counterpart for $p$, namely $S^* = \{r \in I_{p-1}, k_r = k_p \ \mathrm{and} \ t_r = t_p\}$. Using the probabilistic independence as before, we can write:
\begin{eqnarray}
    X_\bk(\bt)
    & = &
    \Ex{\prod_{r \in I_p \setminus S^*} x_{k_r}(t_r)} \, \Ex{\prod_{r \in S^*} x_{k_r}(t_r)} 
    \nonumber \\
    & = &
    \pa{\sum_{\Phi' \in \mathcal{P}(I_p \setminus S^*)} \prod_{S \in \Phi'} \deltabar_{\bk_{S}}(\bt_{S}) \; \mu_{k_{\check{S}}}(t_{\check{S}})}
    \; \deltabar_{\bk_{S^*}}(\bt_{S^*}) \; \mu_{k_{\check{S}^*}}(t_{\check{S}^*})
    \nonumber \\
    & = &
    \sum_{\substack{\Phi = \Phi' \cup \{S^*\} \\ \Phi' \in \mathcal{P}(I_p \setminus S^*)}} \prod_{S \in \Phi} \deltabar_{\bk_{S}}(\bt_{S}) \; \mu_{k_{\check{S}}}(t_{\check{S}})
    \ .
\end{eqnarray}
In the second line, we have used the hypothesis for order $p - |S^*|$ where $|S^*|$ is the number of elements in $S^*$ for the indices that are not in $S^*$, as well as the contraction for all elements in $S^*$ using Eq.~\eqref{eq:dNp}.
The previous expression is valid for each $S^* \subset I_p$ containing $p$, which is determined by $\bk$ and $\bt$.
We conclude by observing that the above dichotomy of partitions $\Phi$ actually spans the whole set $\mathcal{P}(I_p) = \mathcal{P}_p$:
\begin{equation}
    \bigcup_{\substack{S^* \subset I_p \\ S^*\ni p}} \; \bigcup_{\Phi' \in \mathcal{P}(I_p \setminus S^*)} \Phi' \cup \{S^*\}
    =
    \mathcal{P}(I_p) 
    \ ,
\end{equation}
which accounts for all possible configurations of $\bk$ and $\bt$.
This is also related to the decomposition of the Bell number ---giving the number of partitions $\Phi \in \mathcal{P}_p$--- in the sum of the Stirling numbers of the second kind $s_{p,q}$ ---giving the number of partitions $\Phi$ that have $q$ groups. They satisfy the relationship $s_{p,q} = s_{p-1,q-1} + q s_{p-1,q}$ for all $2 \leq q \leq p-1$ (corresponding to the above dichotomy), as well as the ``boundary'' condition $s_{p,q} = 1$ when $q = 1$ or $q = p$.
\end{myproof}
\\
The autocorrelation terms can be represented using diagram representations~\cite{Shchepanyuk_UKJ_1995}.

\subsection{Network with afferent connectivity}

Now that we have introduced definitions and concepts that will be useful to characterize the high-order moments, we turn to the case of a network with afferent connections, but no recurrent connections. The following theorem is the first of our two core results.
We denote the total driving intensity function of the network neurons that lumps together the driving intensity and the input influx by
\begin{equation} \label{eq:driving_rate} 
        \nu_i^{\epsilon = 0}(t)
        =
        \lambda_i(t) + \pa{\gamma_{ik} \ast x_k}(t)
        \ .
\end{equation}

\begin{notation}[Moment for the driving intensities $\lambda$]
To account for possibly stochastic functions $\lambda$ (e.g.\ a Cox process), we define the corresponding moment of order $p$ for the coordinates $\bi \in I_n^p$ at times $\bt = (t_1,\cdots,t_p)$ as 
\begin{equation} \label{eq:mom_lambda}
    \Lambda^p_{\bi}(\bt)
    =
    \Ex{\prod_{r=1}^{p} \lambda_{i_r}(t_r)}_\lambda
    \ .
\end{equation}
\end{notation}

\begin{notation}[Moment for the filtered input]
\label{def:Gamma}
As with $\lambda$, we define the moments of order $p$ of the filtered input $x$ (with afferent kernels $\gamma$) for the coordinates $\bi \in I_n^p$ at times $\bt = (t_1,\cdots,t_p)$ as 
\begin{equation} \label{eq:mom_Gamma}
    \Gamma^p_{\bi}(\bt)
    =
    \Ex{\prod_{r=1}^p\pa{\gamma_{i_rk_r}\ast x_{k_r}}(t_k)}_x
    \ .
\end{equation}
\end{notation}

\begin{definition}[Tensor convolution operator] \label{def:tensor_conv}
Let $\alpha_{ij}: \Re \rightarrow \Re^{n,m}$ be a matrix of kernels. 
We define the $2p$-dimensional tensor that replicates the matrix $\alpha$ for all pairs of indices $(i_r j_r)$:
\begin{equation} \label{eq:tensify_alpha}
    \balpha^p_{\bi\bj}(\bt) = \prod_{r =1}^{p} \alpha_{i_r j_r}(t_r)
\end{equation}
with $\bi = (i_1, \cdots, i_p)\in I_n^p$, $\bj = (j_1, \cdots, j_p)\in I_m^p$ and $\bt = (t_1,\cdots,t_p)$.
For a $p$-order tensor $T^p_{\bj}$ with coordinates $\bj$, the tensor convolution $\circledast$ between $\balpha^p_{\bi \bj}$ and $X_{\bj}$ evaluated at times $\bt$ gives the following tensor of oreder $p$:
\begin{eqnarray} \label{eq:def_tensor_convol}
    \pa{\balpha^p_{\bi\bj} \circledast T^p_\bj}(\bt)
    & = &
    \sum_{\bj = (j_1, \cdots, j_p)} \int_{\bu \in \Re^p} \balpha^p_{\bi \bj}(\bu) T^p_\bj(\bt-\bu) \dd \bu
    \nonumber \\
    & = &
    \pa{ \alpha_{i_1 j_1} \overset{1}{\ast} \cdots \alpha_{i_p j_p} \overset{p}{\ast} T^p_{j_1, \cdots, j_p} }(\bt)
    \ .
\end{eqnarray}
The second line is a reformulation to stress that the convolutions of $\alpha$ are applied on each of the $p$ dimensions ---as indicated above each asterisk--- on the tensor $T^p$, followed by the summation for the tensor product (similar to a matrix product), in line with Eq.~\eqref{eq:def_conv}.
\end{definition}
In essence, this convolution operator involves the same joint ``multiplication'' on paired spatial and temporal dimensions (related to $k_i$ and $t_i$, the temporal convolution being seen as a function multiplication operator) as the matrix convolution in Eq.~\eqref{eq:def_conv}, but extended on all dimensions of the tensor. In particular, this operation is linear.

\begin{property} \label{prop:mom_gamma}
By using the tensor convolution operator defined above, the moments of the filtered inputs can be convenient expressed as
\begin{equation} \label{eq:mom_gamma}
    \Gamma^{p}_{\bi}(\bt)
    =
    \pa{\bgamma^{p}_{\bi \bk} \circledast X^{p}_\bk}(\bt)
    \ ,
\end{equation}
with $\bi = (i_1, \cdots, i_p)\in I_n^p$, $\bk = (k_1, \cdots, k_p)\in I_m^p$ and $\bt = (t_1,\cdots,t_p)$.
\end{property}
\begin{myproof}{Property}{\ref{prop:mom_gamma}}
Let $\bx_\bk^p(\bt) = \prod_{r=1}^p x_{k_r}(t_r)$ be the $p^{\rm th}$ order tensor associated to the input spike train $x$. The moments of the filtered input (see Eq.~(\ref{eq:mom_Gamma})) can be expressed as
\begin{eqnarray}
 \Gamma^{p}_{\bi}(\bt) &=&  \Ex{\prod_{r=1}^p\pa{\gamma_{i_rk_r}\ast x_{k_r}}(t_k)}_x \nonumber\\
&=& \Ex{\pa{\gamma_{i_1k_1} \overset{1}{\ast}\dots\gamma_{i_pk_p} \overset{p}{\ast}\bx^p_{k_1\dots k_p}}(\bt)}_x \nonumber \\
 &=&  \pa{\gamma_{i_1k_1} \overset{1}{\ast}\dots\gamma_{i_pk_p} \overset{p}{\ast}\Ex{\bx^p_{k_1\dots k_p}}_x}(\bt) \nonumber \\
 &=& \pa{\gamma_{\bi\bk} \circledast X^{p}_\bk}(\bt)
\end{eqnarray}
where the last line is obtained from the linearity of the convolution operator and from the definition of the tensor convolution operator defined in Eq.~(\ref{eq:def_tensor_convol}).
\end{myproof}

\begin{theorem}[Input-output mapping for afferent connectivity] \label{th:th_aff}
Consider an uncoupled Hawkes network (definition \ref{def:hawkes}) whose neurons are excited by both inputs $x$ (via afferent connections) and driving intensities $\lambda$, which are probabilistically independent.
The moment $M_{\bi}^{y, \epsilon=0}$ of order $p$ of the network population depends on all smaller-order input moments $X^q$ of the input population as well as moments for the driving intensities $\Lambda^r$ (with $0 \leq q, r \leq p$):
\begin{equation} \label{eq:mom_rec_uncoupled}
    Y_\bi^{p,\epsilon=0}(\bt)
    =
    \sum_{\Phi \in \mathcal{P}_p} \pa{\prod_{S \in \Phi} \deltabar_{\bi_S}(\bt_S)}
    \sum_{\substack{A \cup B = \check{\Phi} \\ A \cap B = \emptyset}} \Gamma^{|A|}_{\bi_A}(\bt_A) \ \Lambda^{|B|}_{\bi_B}(\bt_B)
    \ ,
\end{equation}
where the moments $\Gamma$ and $\Lambda$ are defined in Eqs.~\eqref{eq:mom_lambda} and \eqref{eq:mom_gamma}, respectively.
Here we have defined $\check{\Phi} = \{ \check{S}, S \in \Phi \}$, the set of minima $\check{S} = \min(S)$ over all groups $S$ in the partition $\Phi$.
Note that the superscript of the moment indicates the current situation when the network population is decoupled (i.e.\ $\epsilon = 0$).
\end{theorem}
\begin{myproof}{Theorem}{\ref{th:th_aff}}
Provided the statistics of the inputs $x$ and driving intensities $\lambda$ is known, the spiking activity of the network neurons is determined by the intensity function $\nu_i^{\epsilon = 0}$ in Eq.~\eqref{eq:driving_rate}.
Similar to Eq.~\eqref{eq:prop1_input} in Proposition~\ref{prop:prop_input}, the Poisson nature of the spiking of the network neurons thus gives the following expression for the unconnected neurons with spike trains $y$:
\begin{eqnarray} \label{eq:dev_mom_uncoupled1}
      Y^{p,\epsilon=0}_\bi(\bt)
      & = &
      \Ex{\prod_{r = 1}^{p} y_{i_r}(t_r)}_{y,x,\lambda}
      \nonumber \\
      & = &
      \Ex{\sum_{\Phi \in \mathcal{P}_p} \prod_{S \in \Phi} \deltabar_{\bi_S}(\bt_S) \nu^{\epsilon = 0}_{i_{\check{S}}}(t_{\check{S}})}_{x,\lambda}
      \nonumber \\
      & = &
      \sum_{\Phi \in \mathcal{P}_p} \pa{\prod_{S \in \Phi} \deltabar_{\bi_S}(\bt_S)}
      \Ex{\prod_{S \in \Phi} \nu^{\epsilon = 0}_{i_{\check{S}}}(t_{\check{S}})}_{x,\lambda}
      \nonumber \\
      & = &
      \sum_{\Phi \in \mathcal{P}_p} \pa{\prod_{S \in \Phi} \deltabar_{\bi_S}(\bt_S)}
      \Ex{\prod_{r \in \check{\Phi}} \pa{\lambda_{i_r}(t_r) + \pa{\gamma_{i_r k} \ast x_k}(t_r)}}_{x,\lambda}
      \ .
\end{eqnarray}
In the previous expression, the contractions basically extend the moment of smaller order $|\check{\Phi}| \leq p$ for the intensity $\nu^{\epsilon = 0}$ to the order $p$.
Note that the last line is obtained using the assumption that $\nu^{\epsilon = 0}_i(t) = \nu_i(t) \geq 0$ in Eq.~\eqref{eq:Hawkes} with $\epsilon_{ij} = 0$ for all $i$ and $t$.

The product involving the sum of $\lambda_{i_{\check{S}}} + \gamma_{i_{\check{S}} k} \ast x_k$ gives $2^{|\check{\Phi}|}$ terms with $|\check{\Phi}|$ being the number of elements in $\check{\Phi}$. 
Now we develop this product to isolate the contributions originating from the input moments of the same order on the one hand, and from the driving intensities on the other hand, using the fact that they are statistically independent.
To this end, we use the following expression that converts a product of a sum into a sum of products:
\begin{equation}
    \prod_{r \in C}(a_r + b_r) 
    =
    \sum_{\substack{A \cup B = C \\ A \cap B = \emptyset}} \left( \prod_{r \in A} a_r \right)\left( \prod_{rÕ \in B} b_{rÕ} \right)
    \ ,
\end{equation}
where $A$ and $B$ can be empty sets.
In our case, $A \subset I_p$ is the subset of indices belonging to $\check{\Phi}$ that concern input neurons in Eq.~\eqref{eq:dev_mom_uncoupled1}, while $B = \check{\Phi} \setminus A$ is the subset of indices that concern $\lambda$.
Because the random variables $x$ and $\lambda$ are independent, this gives
\begin{eqnarray} \label{eq:res_th1}
      Y^{p,\epsilon=0}_\bi(\bt)
      & = &
      \sum_{\Phi \in \mathcal{P}_p} \pa{\prod_{S \in \Phi} \deltabar_{\bi_S}(\bt_S)}
      \sum_{\substack{A \cup B = \check{\Phi} \\ A \cap B = \emptyset}} 
      \Ex{\prod_{r \in A} \pa{\gamma_{i_r k}\ast x_k}(t_r)}_x \, \Ex{\prod_{r' \in B} \lambda_{i_{r'}}(t_{r'})}_\lambda
      \nonumber \\
      & = &
      \sum_{\Phi \in \mathcal{P}_p} \pa{\prod_{S \in \Phi} \deltabar_{\bi_S}(\bt_S)}
      \sum_{\substack{A \cup B = \check{\Phi} \\ A \cap B = \emptyset}} \Gamma^{|A|}_{\bi_A}(\bt_A) \ \Lambda^{|B|}_{\bi_B}(\bt_B)
      \ .
\end{eqnarray}
\end{myproof}
\\
Note that the expression in Eq.~\eqref{eq:res_th1} can be rewritten by grouping together the moments of order $|A| = q$ and $|B| = r$ under the form
\begin{equation}
    Y_\bi^{p,\epsilon=0}(\bt)
    =
    \sum_{0 \leq q + r \leq p} \ \mathcal{A}^p\pasq{\Gamma^{q}, \Lambda^{r} }_{\bi}(\bt)
\end{equation}
where $\mathcal{A}^p$ is an operator that considers all possible combinations with the delta functions, see Appendix~\ref{an:op_aff}.

\begin{remark}
When the network of unconnected neurons is not driven by an external intensity ($\lambda = 0$), Eq.~\eqref{eq:res_th1} can be simplified as
\begin{equation} \label{eq:mom_no_lambda}
    Y_{\bi}^{p,\epsilon=0}(\bt)
    =
    \sum_{\Phi \in \mathcal{P}_p} \pa{\prod_{S \in \Phi} \deltabar_{\bi_S}(\bt_S)}\Gamma_{\bi_{\check{\Phi}}}^{|\check{\Phi}|}(t_{\check{\Phi}})
\end{equation}
Even in this general case, the output moment $Y^{p,\epsilon=0}$ of order $p$ is an intricate function of the moments of orders $r\leq p$, i.e.\ it depends on $X^r$ via $\Gamma^r$ with $r = |\check{\Phi}|$.
This contrasts with the fact that the moment $\Gamma^{p}$ only depends on the corresponding moment $X^{p}$ of the same order, see Property~\ref{prop:mom_gamma}.
\\
Conversely, in the absence of spiking inputs ($\gamma = 0$) and when the driving intensities $\lambda_i(t)$ are deterministic, the moments $\Lambda$ simply come from the multiplication of the intensity functions:
\begin{equation}
    Y_{\bi}^{p,\epsilon=0}(\bt)
    =
    \sum_{\Phi \in \mathcal{P}_p} \prod_{S \in \Phi} \pa{ \deltabar_{\bi_S}(\bt_S)\lambda_{\bi_{\check{S}}}(\bt_{\check{S}})}
    \ .
\end{equation}
\end{remark}

\subsection{Network with recurrent connectivity}

The last step is to consider connections determined by $\epsilon$ between the network neurons, the second half of our core result. 

\begin{definition}[Effective recurrent kernel]
Let $\tepsilon: \Re \rightarrow \Re^{n\times n}$ denote the effective recurrent kernel and be defined as
\begin{equation} \label{eq:rec_kernel}
    \tepsilon_{ij}(t) = \sum_{n\geq 0}\epsilon_{ij}^{\ast n}(t)
\end{equation}
where 
\begin{equation}
    \epsilon_{ij}^{\ast n}(t) =
    \left\{\begin{array}{ll} 
    \Big(\epsilon_{il}^{\ast (n-1)} \ast \epsilon_{lj}\Big)(t) & \mathrm{if} \quad n>0
    \\
    \deltabar_{ij}(t,0) & \mathrm{if} \quad n=0
    \end{array}\right.
\end{equation}
is the $n^\mathrm{th}$ order convolution. 
\end{definition}
Recall that the convolution is defined for kernel matrices, see Eq.~\eqref{eq:def_conv}. Because $\epsilon_{ij}(t) = 0$ for $t \leq 0$ and all pairs $(i,j)$ (due to the causality requirement), $\tepsilon_{ij}(t) = 0$ as well for $t \leq 0$.
This effective recurrent kernel is the equivalent in the time domain to the matrix inverse of the identity minus the ``spatio-temporal'' connectivity in the Fourier domain~\cite{Hawkes_JRSB_1971}.

\begin{property} \label{prop:inv_eps}
The effective recurrent kernel $\tepsilon$ satisfies the following self-consistency equation:
\begin{equation} \label{eq:prop_inv_eps}
    \pa{ \kappa \ast \tepsilon }_{ij}(t)
    =
    \deltabar_{ij}(t,0)
    \ ,
\end{equation}
where $\kappa_{ij}(t) = \deltabar_{ij}(t,0) - \epsilon_{ij}(t)$.
Therefore, $\tepsilon$ can be thought as the inverse of $\kappa$ for the convolution operator.
\end{property}
\begin{myproof}{Property}{\ref{prop:inv_eps}}
By convolving the $\epsilon$ kernel with the effective recurrent kernel $\tepsilon$, we find (omitting the time variables)
\begin{equation}
    \pa{\epsilon\ast\tepsilon}_{ij}
    =
    \epsilon_{ik}\ast\pa{\sum_{n \geq 0}\epsilon_{kj}^{\ast n}}
    =
    \sum_{n \geq 1}\epsilon_{ij}^{\ast n} = \tepsilon_{ij} - \deltabar_{ij}
    \ ,
\end{equation}
which we reorganize to factorize $\tepsilon$, obtaining Eq.~\eqref{eq:prop_inv_eps}.
\end{myproof}

\begin{example}[Single neuron with self-connection and with driving intensity $\lambda$] \label{ex:ex_rec}

We firstly present an illustrative version of our proof by induction for a single neuron with self-feedback and driven by a deterministic intensity $\lambda$ in the cases $1 \leq p \leq 3$. In this example $\Ex{\cdots} = \Ex{\cdots}_y$ as there is no other source of stochasticity. Note that $p = 2$ corresponds to Hawkes' results~\cite{Hawkes_JRSB_1971} with moments instead of (auto)covariances. The motivation is providing a concrete case for stepping from orders $p$ to $p + 1$, which is formalized in the proof below.

\underline{Cases $p = 1$ and $p = 2$:}

The first-order moment for $p = 1$ corresponds to the mean firing rate and can be calculated from the driving intensity function $\lambda$ by solving the self-consistency equation given by the second line of Eq.~\eqref{eq:Hawkes} using the equality for the intensity $\langle y(t) \rangle = \langle \nu(t) \rangle$:
\begin{equation} \label{eq:sol_1st_order}
    \Ex{y(t)} = \pa{\tepsilon \ast \lambda}(t)
    \ .
\end{equation}

For the second order, the point is to take into account the effects of spikes upon the future spiking probability, with the effect of the self-feedback loop.
Assuming $t_1 \leq t_2$ (gray semi-plane in Fig.~\ref{fig3_rec_single_th}A), we can develop $y(t_2)$ in $\langle y y \rangle (t_1, t_2)$ using Eq.~\eqref{eq:Hawkes}. This holds because the intensity function $\nu(t_2)$ requires the knowledge of past spiking activity $y(u)$ with $u < t_2$, as illustrated by the dark gray arrow in Fig.~\ref{fig3_rec_single_th}A, moving toward the diagonal $t_1 = t_2$.
This development gives
\begin{eqnarray} \label{eq:req_2nd_order}
    \langle y y \rangle (t_1, t_2)
    & = &
    \langle y(t_1) \nu(t_2) \rangle + \langle y(t_1) \rangle \delta(t_2 - t_1)
    \nonumber \\
    & = &
    \langle y(t_1) \big(\epsilon \ast y\big)(t_2) \rangle + \langle y(t_1) \rangle \lambda(t_2) + \langle y(t_1) \rangle \delta(t_2 - t_1)
    \nonumber \\
    & = &
    \big( \epsilon \overset{2}{\ast} \overline{\langle y y \rangle} \big)(t_1, t_2) + \overline{\langle y \rangle \lambda}(t_1, t_2) + \overline{\langle y \rangle \delta^{21}}(t_1, t_2)
    \ .
\end{eqnarray}
Note that $\nu$ is inside the angular brackets on the right-hand side of the first line, because $\nu(t_2)$ and $y(t_1)$ are not independent when the difference in the time variables lies within the range of $\tepsilon$.
The last line is simply a rewriting using a specific notation with a line above multivariate functions to indicate the order of the functions with respect to the time variables, which will be useful for this example. In addition, we use the notation introduce in Eq.~\eqref{eq:def_tensor_convol} where $\overset{2}{\ast}$ indicates the convolution performed on the second time variable $t_2$ and the Dirac delta $\delta^{21}(t_2) := \delta(t_2 - t_1)$ is a redundant expression as a function of $t_2$, while keeping the information about $t_1$. 

The solution $\langle y y \rangle (t_1, t_2)$ must satisfy Eq.~\eqref{eq:req_2nd_order} for all $t_1 \leq t_2$, which is a Wiener-Hopf equation. The atomic contribution (Dirac delta) acts as a ``boundary condition'' when $t_2 \rightarrow t_1$.
Our strategy is the following: we propose a solution for the moment of order $p = 2$ and verify that it satisfies the required Eq.~\eqref{eq:req_2nd_order}. As the solution is fully symmetric in $t_1$ and $t_2$, this implies that the solution is also valid on the complementary space $t_2 \leq t_1$, being eventually valid for all $(t_1, t_2) \in \Re^2$.
The putative 2nd-order moment is:
\begin{equation} \label{eq:sol_2nd_order}
    \overline{\langle y y \rangle} (t_1, t_2)
    =
    \pa{ \tepsilon \overset{1}{\ast} \tepsilon \overset{2}{\ast} (\overline{\lambda \lambda} + \overline{\lambda \delta^{21}}) }(t_1, t_2)
    \ , 
\end{equation} 
Note that our notation does not require the time variables, allowing for compact writing.
We use the equality in Eq.~\eqref{eq:prop_inv_eps} on $\tepsilon \overset{2}{\ast}$ to obtain
\begin{eqnarray} \label{eq:2nd_order_dev}
    \overline{\langle y y \rangle}
    & = &
    \tepsilon \overset{1}{\ast} (\epsilon \ast \tepsilon + \delta) \overset{2}{\ast} (\overline{\lambda \lambda} + \overline{\lambda \delta^{21}})
    \nonumber \\
    & = &
    \epsilon \overset{2}{\ast} \overline{\langle y y \rangle} + \tepsilon \overset{1}{\ast} \overline{\lambda \lambda} + \tepsilon \overset{1}{\ast} \overline{\lambda \delta^{21}}
    \ . 
\end{eqnarray} 
For the first term of the right-hand side in the upper line, the convolution by $\epsilon \ast \tepsilon$ on the second variable $t_2$ has been rewritten by moving $\epsilon$ out, while the rest is in fact $\overline{\langle y y \rangle}$ in Eq.~\eqref{eq:sol_2nd_order}. In the second term, the convolution by the Dirac on $t_2$ and we obtain two terms involving $\tepsilon \ast \lambda (t_1) = \langle y \rangle (t_1)$, see the solution for the 1st-order moment in Eq.~\eqref{eq:sol_1st_order}. Together, these three terms are the right-hand side of Eq.~\eqref{eq:req_2nd_order}, which is thus satisfied.

Note also that $\tepsilon(t) = 0$ for $t < 0$ (reflecting causality of the overall ``feedback' kernel), which implies that the operator $\tepsilon \overset{1}{\ast} \tepsilon \overset{2}{\ast}$ applied on the 2-dimensional function under the overline only ``spreads'' the function mass towards future (see Fig.~\ref{fig3_rec_single_th}B).

\begin{figure}
        \includegraphics[scale=1]{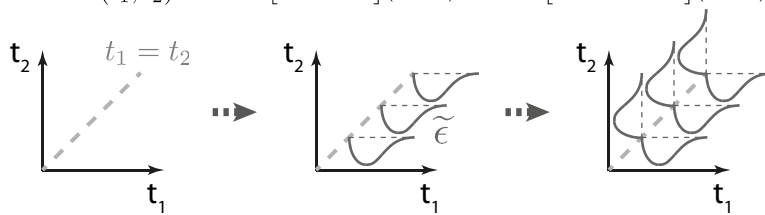}
        \caption{ \label{fig3_rec_single_th}
        {\bf Schematic diagrams supporting the calculations for the 2nd- and 3rd-order moments.}
        {\bf A:} The development in Eq.~\eqref{eq:req_2nd_order} corresponds in expressing $y(t_2)$ as a function of the past history. This requires that $t_2 > t_1$, as illustrated by the gray upper triangle of the plane. The dark-gray arrow indicates the ``direction'' of the development towards the past network activity (related to the convolution by $\epsilon$), which is necessary to evaluate the firing probabilities involved in the moment.
        {\bf B:} Schematic representation of the twofold convolution involved in Eq.~\eqref{eq:sol_2nd_order} for the calculation of the second-order moment. The Dirac delta correspond to a function that is non-zero on the diagonal $t_1 = t_2$ only, as represented by the gray dashed line. The effect of the first convolution on $t_1$ ``spreads'' the diagonal function towards the ``future'' in the horizontal direction. Then, the convolution on $t_2$ ``spreads'' the whole towards the ``future'' in the vertical direction, resulting in a symmetric function. Note that the result is distinct from outer product of the time vectors $\overline{(\tepsilon \ast \lambda)(\tepsilon \ast \lambda)}$.
        {\bf C:} Similar diagram to panel A to indicate the subspace for the condition $t_1 \leq t_2 \leq t_3$ and represent the development of the moment for $p = 3$ in Eq.~\eqref{eq:req_3rd_order}.
        }
\end{figure}

\underline{Case $p = 3$:}

Following the previous section, we extend the calculations to the case $p = 3$ in order to prepare for the generalization to arbitrary $p \geq 2$. As with $p= 2$, we consider the ordering $t_1 \leq t_2 \leq t_3$ (gray subspace in Fig.~\ref{fig3_rec_single_th}C), which allows the development of the third time variable as was done in Eq.~\eqref{eq:req_2nd_order}
\begin{equation} \label{eq:req_3rd_order}
    \overline{\langle y y y \rangle} (t_1, t_2, t_3)
    =
    \epsilon \overset{3}{\ast} \overline{\langle y y y \rangle} (t_1, t_2, t_3)
    + 
    \overline{\langle y y \rangle \lambda} (t_1, t_2, t_3)
    +
    \overline{\langle y y \rangle \delta^{32}} (t_1, t_2, t_3)
    \ ,
\end{equation}
with the Dirac corresponding to the ``boundary condition'' when $t_3 \rightarrow t_2$, corresponding to the ``lower'' tilted plane of the gray subspace to which points the dark gray arrow in Fig.~\ref{fig3_rec_single_th}C.
Note that this involves only the atomic contribution $\delta^{32}$ ($\delta^{21}$ is in $y y$ corresponding to $(t_1, t_2)$), the other $\delta^{31}$ alone is not possible in this space. See also the discussion in Example~\ref{ex:ex_in} for the second-order input moments.
Now we pursue the calculations without the time variables in arguments, as before for $p = 2$.
The putative symmetric solution is
\begin{equation}
    \overline{\langle y y y \rangle}
    =
    \tepsilon \overset{1}{\ast} \tepsilon \overset{2}{\ast} \tepsilon \overset{3}{\ast} (\overline{\lambda \lambda \lambda} + \overline{\lambda \lambda \delta^{32}} + \overline{\lambda \lambda \delta^{31}} + \overline{\lambda \delta^{21} \lambda} + \overline{\lambda \delta^{21} \delta^{32}})
    \ ,
\end{equation}
which involves the contractions for all partitions of $\{1, 2, 3\}$, in a similar fashion to Eq.~\eqref{eq:mom_rec_uncoupled}.
We use again Eq.~\eqref{eq:prop_inv_eps} as in Eq.~\eqref{eq:2nd_order_dev} to obtain the convolution of $\epsilon$ with $\overline{\langle y y y \rangle}$ on $t_3$ and regroup the other terms where the convolution with $t_3$ vanishes because of the Dirac in order to use the expression of the 2nd-order moment in Eq.~\eqref{eq:sol_2nd_order}, namely $\tepsilon \overset{1}{\ast} \tepsilon \overset{2}{\ast} (\overline{\lambda \lambda} + \overline{\lambda \delta^{21}}) = \overline{\langle y y \rangle}$:
\begin{eqnarray} \label{eq:3rd_order_dev}
    \overline{\langle y y y \rangle}
    & = &
    \epsilon \overset{3}{\ast} \overline{\langle y y y \rangle}
    + 
    \tepsilon \overset{1}{\ast} \tepsilon \overset{2}{\ast} (\overline{\lambda \lambda \lambda} + \overline{\lambda \delta^{21} \lambda})
    + 
    \tepsilon \overset{1}{\ast} \tepsilon \overset{2}{\ast} (\overline{\lambda \lambda \delta^{32}} + \overline{\lambda \delta^{21} \delta^{32}})
    + 
    \tepsilon \overset{1}{\ast} \tepsilon \overset{2}{\ast} \overline{\lambda \lambda \delta^{31}}
    \nonumber \\
    & = &
    \epsilon \overset{3}{\ast} \overline{\langle y y y \rangle}
    +
    \overline{\langle y y \rangle \lambda}
    +
    \overline{\langle y y \rangle \delta^{32}}
    + 
    \tepsilon \overset{1}{\ast} \tepsilon \overset{2}{\ast} \overline{\lambda \lambda \delta^{31}}
    \ .
\end{eqnarray}
What remains to be seen is that the condition $t_1 \leq t_2 \leq t_3$ implies that $\delta^{31} = 0$ always: when $t_1 = t_3$, in fact we have $t_1 = t_2 = t_3$, which corresponds to $\delta^{21} \delta^{31}$.
This means that the last term in Eq.~\eqref{eq:3rd_order_dev} vanishes and Eq.~\eqref{eq:req_3rd_order} is satisfied. The symmetry argument ensures the validity over all $(t_1, t_2, t_3)$, as will be formalized below.

\underline{Numerical simulation:}

\begin{figure}
        	\includegraphics[scale=0.9]{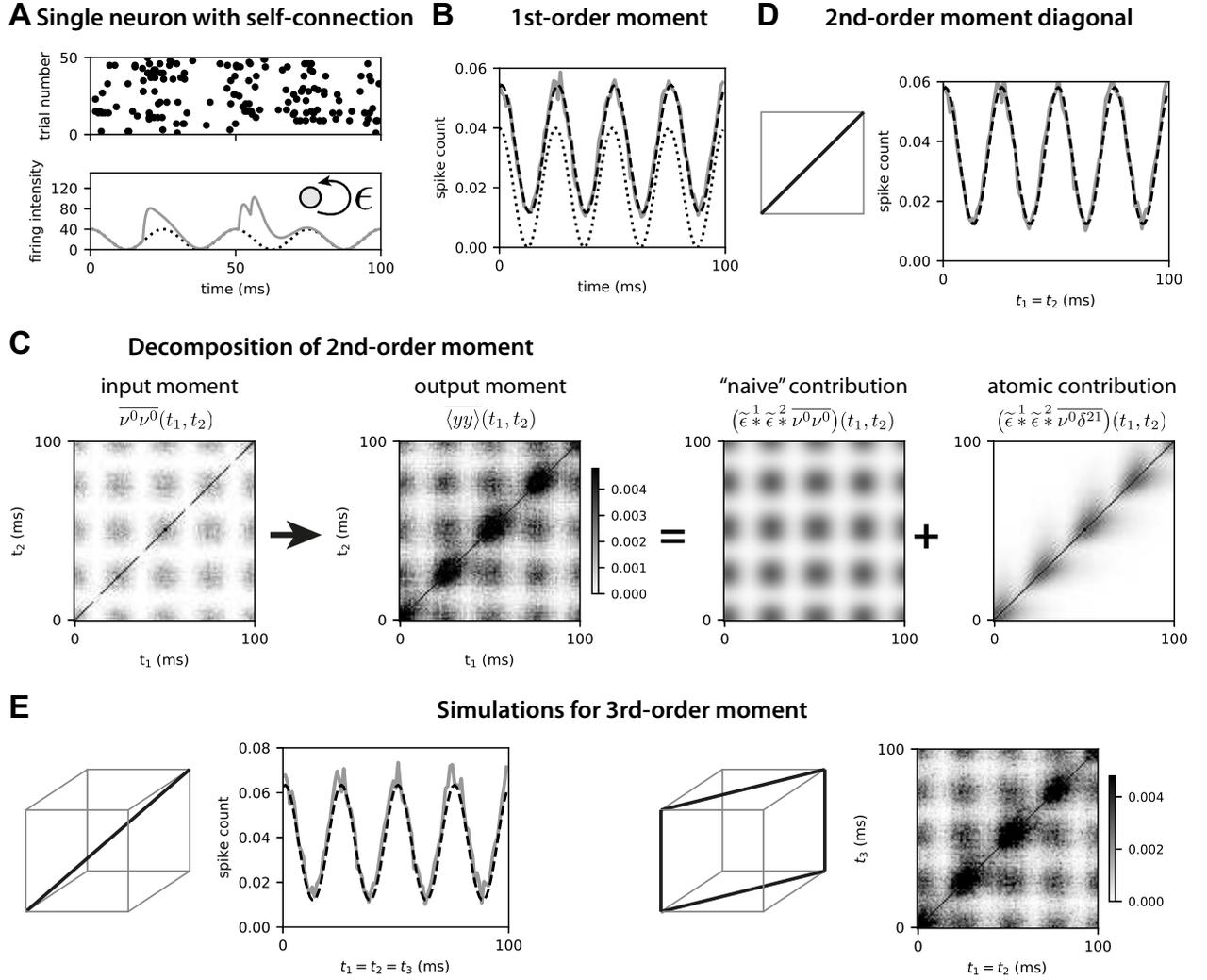}
        	\caption{ \label{fig4_rec_single_sim}
        	{\bf Output moments for a single neuron with self-connection.}
        {\bf A:} Spike raster (top plot) for 50 simulations for a neuron, similar to Fig.~\ref{fig2_mom_cmplt}.
        In the bottom plot, the driving oscillatory intensity $\lambda$ (dotted black curve) is compared with the firing intensity $\nu$ (green curve), which is affected by the neuron's firing.
        {\bf B:} 1st-order moment (solid gray curve) with theoretical prediction (dashed black curve).
        The dotted black curve indicate the driving intensity $\lambda$.
        {\bf C:} The two left plots represent the Input-output mapping for the 2nd-order moment, averaged over 10000 simulations (darker pixels indicate a higher spike density). 
        The two right plots illustrate the decomposition into a contribution due to rate correlation (co-fluctuations, ``naive'' contribution) and that due to autocorrelation. Note that the right plot corresponds to Fig.~\ref{fig3_rec_single_th}B.
        The equations above refer to the terms in Eq.~\eqref{eq:2nd_order_dev}.
        {\bf D:} Simulation (gray curve) and theoretical prediction (dashed black curve) of the diagonal of the matrix for the output moment in panel C.
        {\bf E:} Example slices for the 3rd-order moment, as indicated by the left diagram (color coded).
        All prediction curves are calculated using Eqs.~\eqref{eq:2nd_order_dev} and \eqref{eq:3rd_order_dev}.
        	}
\end{figure}

The upper plot in Fig.~\ref{fig4_rec_single_sim}A illustrates that the rhythm of the output spiking is altered by the recurrent self-connection.
This comes from the fact that, for an excitatory self-connection, output spikes momentarily increase the firing probability, as can be seen when comparing the green curve with the dotted black curve in the bottom plot. 
The output first-order moment in Fig.~\ref{fig4_rec_single_sim}B (solid gray curve for the simulation and dashed black curve for the prediction) is above the input first-order moment related to the underlying driving intensity $\lambda$ (dotted black curve). Note also the shift to later time.

The decomposition of the second-order moment in Fig.~\ref{fig4_rec_single_sim}C illustrates that the effect of autocorrelations (right plot) spreads from the diagonal due to the self-connection. The main diagonal for $p = 2$ in Fig.~\ref{fig4_rec_single_sim}D has larger values than the curve for $p = 1$ in Fig.~\ref{fig4_rec_single_sim}B. In Fig.~\ref{fig4_rec_single_sim}E, the main diagonal for $p = 3$ (gray curve in the left plot) is even larger, indicating that effects due to autocorrelation cumulate (as for input moments in Fig.~\ref{fig2_mom_cmplt}).
The slice of the output third-order moment (right matrix in Fig.~\ref{fig4_rec_single_sim}E) has smaller value, but note the high spike density along the diagonal of the right matrix due to the spreading of atomic contributions by the recurrent kernel $\epsilon$.

\end{example}

\begin{theorem}[Input-output mapping for recurrent connectivity] \label{th:th_rec}
The moment $Y^p_{\bi}$ of order $p$ of the Hawkes process (definition \ref{def:hawkes}) of the network population can be expressed as
\begin{equation} \label{eq:net_mapping}
    Y^p_{\bi}(\bt)
    =
    \pa{\btepsilon^p_{\bi\bj} \circledast Y^{p,\epsilon=0}_\bj}(\bt)
    \ .
\end{equation}
The effects of the recurrent connectivity on the input moments are determined by spatio-temporal filtering described by the effective recurrent kernel $\btepsilon^p$ defined similarly to Eq.~\eqref{eq:tensify_alpha} on the moment for uncoupled neurons in Eq.~\eqref{eq:mom_rec_uncoupled}.
\end{theorem}
%

\begin{myproof}{Theorem}{\ref{th:th_rec}} \label{proof:proof2} 

Compared to Example~\ref{ex:ex_rec}, we consider the general case where inputs and/or external intensities drive the network neurons via $\nu^{\epsilon = 0}$ in Eq.~\eqref{eq:driving_rate}. Let introduce the \emph{conditional moment} $M^p_{\bi}(\bt)$ of order $p$ defined as
\begin{equation} \label{eq:def_M_p}
    M^p_{\bi}(\bt)
    =
    \Ex{ \prod_{r=1}^p y_{i_r}(t_r)}_{y|x,\lambda}
    \ ,
\end{equation}
where the conditioning is over the input activity $x$ and the driving intensities $\lambda$.
Note that the statistical averaging over $x$ and $\lambda$ of the conditional moment gives the (unconditional) moment defined in Eq.~\eqref{eq:mom_out}: $\Ex{M_\bi^p(\bt)}_{x,\lambda} = Y_\bi^p(\bt)$.
To demonstrate Eq.~\eqref{eq:net_mapping}, we prove by induction the following result on $M_\bi^p(\bt)$, which straightforwardly leads to the expression in Theorem~\ref{th:th_rec} by taking the same statistical averaging over $x$ and $\lambda$ as done above:
\begin{equation}\label{eq:sol_order_p}
  M^p_{\bi}(\bt)
  =
  \pa{\btepsilon^p_{\bi\bj} \circledast M^{p,\epsilon=0}_\bj}(\bt)
  \ ,
\end{equation}
where the conditional moment of order $p$ in the absence of recurrent coupling ($\epsilon = 0$) is defined as
\begin{equation}
    M^{p,\epsilon=0}_{\bj}(\bt)
    =
    \sum_{\Phi \in \mathcal{P}_{p}} \prod_{S \in \Phi} \deltabar_{\bj_S}(\bt_S) \ \nu^{\epsilon = 0}_{j_{\check{S}}}\!\pa{t_{\check{S}}}
\end{equation}
In Eq.~\eqref{eq:sol_order_p} the effect of the past spiking activity of $y$ due to the recurrent connectivity $\epsilon$ is taken care of by all $\tepsilon$, considering $\nu^{\epsilon = 0}$ to be ``deterministic'' from the viewpoint of $y$ provided $x$ and $\lambda$ are known.

The \emph{conditioned moment} $M^p_{\bi}(\bt)$ in Eq.~\eqref{eq:sol_order_p} must obey the constraints imposed by the dynamics in Eq.~\eqref{eq:Hawkes}.
Under the condition on the time variables $t_1 \leq \cdots \leq t_p$, we can develop for $y_{i_p}(t_p)$ using the past activity of $(y_1(t), \cdots,y_p(t))$ for $t < t_p$ and the intensity $\nu^{\epsilon=0}_{i_p}(t_p)$. Let $\bi = (i_1,\cdots,i_p)$ denote the coordinates and $\bt = (t_1,\cdots,t_p)$ the time variables. The $p^\mathrm{th}$ order correlation of the output population can be expressed as
\begin{eqnarray} \label{eq:req_order_p}
    M^p_{\bi}(\bt)
    & = &
    \Ex{\prod_{r=1}^{p}y_{i_r}(t_r)}_{y|x,\lambda}
    \nonumber \\
    &=&
    \Ex{ \prod_{r=1}^{p-1} y_{i_r}(t_r) \cdot \nu_{i_p}(t_p) }_{y|x,\lambda} 
    + \Ex{ \prod_{r=1}^{p-1} y_{i_r}(t_r) }_{y|x,\lambda} \; \deltabar_{i_{p-1}i_p}(t_{p-1},t_p)
    \ .
\end{eqnarray}
Note that the generalized delta corresponds to the ``boundary condition'' $t_{p} = t_{p-1}$, as done in the above examples to moments. A similar condition for the time lag was used in the case of covariances~\cite{Hawkes_JRSB_1971,Gilson_BC_2009b}. By using the development of $\nu_i(t) = \pa{ \epsilon_{ij} \ast y_j }(t) + \nu_i^{\epsilon=0}(t)$, see Eqs.~\eqref{eq:Hawkes} and~\eqref{eq:driving_rate} and by setting  $\bi' = (i_1,\cdots,i_{p-1})$ which contains the $p-1$ first elements of $\bi$, and similarly $\bt' = (t_1,\cdots,t_{p-1})$, we have
\begin{eqnarray} \label{eq:req2_order_p}
     M^p_{\bi}(\bt)
     & = &
    \pa{\epsilon_{i_p,j_p} \overset{p}{\ast} M^p_{\bi'j_p}}(\bt) 
    + \Ex{ \prod_{r=1}^{p-1} y_{i_r}(t_r) \cdot \nu_{i_p}^{\epsilon=0}(t_p) }_{y|x,\lambda} 
    + \Ex{ \prod_{r=1}^{p-1} y_{i_r}(t_r) }_{y|x,\lambda}  \; \deltabar_{i_{p-1}i_p}(t_{p-1},t_p)
    \nonumber \\
    &=&
    \pa{\epsilon_{i_p,j_p} \overset{p}{\ast} M^p_{\bi'j_p}}(\bt) 
    + M^{p-1}_{\bi'}(\bt') \; \pa{ \nu_{i_p}^{\epsilon=0}(t_p) + \deltabar_{i_{p-1}i_p}(t_{p-1},t_p) } 
     \ .
\end{eqnarray}
where the conditioned moment of order $p-1$ appears in the right-hand side. Therefore, we can use Eq.~\eqref{eq:sol_order_p} for the order $p-1$:
\begin{eqnarray} \label{eq:dev_order_p}
    \lefteqn{
    M^{p-1}_{\bi'}(\bt') \; \pa{ \nu_{i_p}^{\epsilon=0}(t_p) + \deltabar_{i_{p-1}i_p}(t_{p-1},t_p) }
    }
    \nonumber \\
    & = &
    \pa{\btepsilon^{p-1}_{\bi'\bj'} \circledast M^{p-1,\epsilon=0}_{\bj'}}(\bt') \; 
    \pa{ \nu_{i_p}^{\epsilon=0}(t_p) + \deltabar_{i_{p-1}i_p}(t_{p-1},t_p) }
    \nonumber \\
    & = &
    \btepsilon^{p-1}_{\bi'\bj'} \overset{p-1}{\circledast}
    \pa{ \sum_{\Phi \in \mathcal{P}^0_{p-1}} \prod_{S \in \Phi} \deltabar_{\bj_S}(\bt_S) \ \nu^{\epsilon = 0}_{j_{\check{S}}}\!\pa{t_{\check{S}}} 
    \pa{ \nu_{i_p}^{\epsilon=0}(t_p) + \deltabar_{i_{p-1}i_p}(t_{p-1},t_p) } }
    \nonumber \\
    & = &
    \btepsilon^{p-1}_{\bi'\bj'} \overset{p-1}{\circledast}
    \pa{ \sum_{\Phi \in \mathcal{P}^0_{p}} \prod_{S \in \Phi} \deltabar_{\bj_S}(\bt_S) \ \nu^{\epsilon = 0}_{j_{\check{S}}}\!\pa{t_{\check{S}}} }
    \nonumber \\
    & = &
    \btepsilon^{p-1}_{\bi'\bj'} \overset{p-1}{\circledast} M^{p,\epsilon=0}_{\bj' i_p}(\bt)
    \ .
\end{eqnarray}
Note that the tensor convolution $\overset{p-1}{\circledast}$ applies to the first $p-1$ indices $j' = (j_1, \cdots, j_{p-1})$ of the tensor of dimension $p$.
In the third line of Eq.~\eqref{eq:dev_order_p}, we only retain the partitions that contribute to the summation under the condition $t_1 \leq \cdots \leq t_p$.
To do so we define the subset $\mathcal{P}^0_{p-1} \subset \mathcal{P}_{p-1}$ of ordered partitions $\Phi$, where the groups $S \in \Phi$ consist of all successive indices between $\check{S} = \min(S)$ and $\max(S)$ (equal for singletons).
Following, we integrate the elements in the squared brackets to the sum by augmenting the partitions $\Phi \in \mathcal{P}^0_{p-1}$ to partitions in $\mathcal{P}^0_{p}$.
Note that the passage from the second line to the fifth line in Eq.~\eqref{eq:dev_order_p} also corresponds to taking $\epsilon = 0$ in Eq.~\eqref{eq:req_order_p}.

Going back to Eq.~\eqref{eq:req2_order_p}, we isolate $M^p_{\bi}(\bt)$ on the left-hand side:
\begin{equation} \label{eq:dev2_order_p}
    \pa{ \kappa_{i_p,j_p} \overset{p}{\ast} M_{\bi' j_p}^p }(\bt)
    = 
    \btepsilon^{p-1}_{\bi'\bj'} \overset{p-1}{\circledast} M^{p,\epsilon=0}_{\bj' i_p}(\bt)
    \ ,
\end{equation}
where $\kappa_{ij}(t) = \deltabar_{ij}(t,0) - \epsilon_{ij}(t)$ (see Prop~\ref{prop:inv_eps}).
Using the property of $\tepsilon$ in Eq.~\eqref{eq:prop_inv_eps}, we obtain
\begin{eqnarray} \label{eq:dev2_order_p}
    M^p_{\bi}(\bt)
    & = &
    \pa{ \tepsilon_{i_p j'_p} \overset{p}{\ast} \pa{ \kappa_{j'_p,j_p} \overset{p}{\ast} M_{\bi' j_p}^p } }(\bt)
    \nonumber \\
    &=&
    \pa{\tepsilon_{i_p j_p} \overset{p}{\ast}    \pa{\btepsilon^{p-1}_{\bi'\bj'} \overset{p-1}{\circledast} M^{p,\epsilon=0}_{\bj'j_p}}} (\bt)
    \nonumber \\
    &=&
    \pa{\btepsilon^{p}_{\bi\bj} \circledast M^{p,\epsilon=0}_{\bj}} (\bt)
    \ .
\end{eqnarray}

Note so far we have only established the validity of this result for $t_1\leq\cdots\leq t_p$. As said above, this is equivalent of considering only ordered partitions $\mathcal{P}^0_p$. The generalization to an arbitrary $\bt = (t_1, \cdots, t_p)$ can be obtained by noting that an arbitrary $\bt = (t_1, \cdots, t_p)$ can be mapped to an ordered version using permutations, say $\Pi(\bt) = \bt' = (t'_1 \leq \cdots \leq t'_p)$.
The partition set $\mathcal{P}^0_p$ is thus replaced by $\pacu{ \Pi(\Phi), \Phi \in \mathcal{P}^0_p}$ with $\Pi(\Phi)$ being the partition of the image indices via $\Pi$.
Note that this covers entire set of all partitions $\mathcal{P}_p$ when considering all possible permutations.
This concludes the proof by induction.

\end{myproof}
\begin{remark}[Large population size] \label{rem:approx}
In the limit of large population size ($n \rightarrow \infty$) and in the absence of the driving intensities ($\lambda = 0$), the output moment of order $p$ can simply be approximated by the single dominating term
\begin{equation}
    Y_{\bi}^p(\bt) \simeq \pa{\btepsilon^p_{\bi \bj} \circledast \bgamma^p_{\bj\bk}\circledast X^p_{\bk}}(\bt)
\end{equation}
This corresponds to the partition $\Phi = \pacu{I_p}$ and has a contribution of order $n^p$ whereas all other partitions $\Phi'\neq\Phi$ give a contribution of order $n^{p-|\Phi'|+1}\ll n^p$ which is negligible.
\end{remark}

\subsection{Further examples}

\begin{example}[Interplay between afferent and recurrent connections]

Here we consider three cases of a single neuron where the amplification determined by the weights is the same, namely 
the integral of $\gamma \ast \tepsilon$ is identical across the three cases.
In this way, the output neuron has the same firing rate in all cases and the point is the comparison of its spike-correlation structure. We rescale unitary kernels rescaled by the following weights $w_\mathrm{aff}$ and $w_\mathrm{rec}$ for $\gamma$ and $\epsilon$, respectively. The simulation results in Fig.~\ref{fig5_aff_rec} show that the distinct types of connectivity have strong influences on the output correlation structure. The combination of afferent and recurrent connections leads to a spreading of the density of the 2nd- and third-order moments. Although the afferent connection does not amplify the driving input in the right configuration, the autocorrelation of the input neuron contributes to a stronger correlation structure for the output neuron. Recall that, because all configurations have the same firing rate, the difference lies in the temporal distribution of the spikes, which are more bursty due to the recurrent connectivity. In other words, the presence of the input neuron with the afferent connection further strengthens the bursting.

\begin{figure}
        	\includegraphics[scale=0.95]{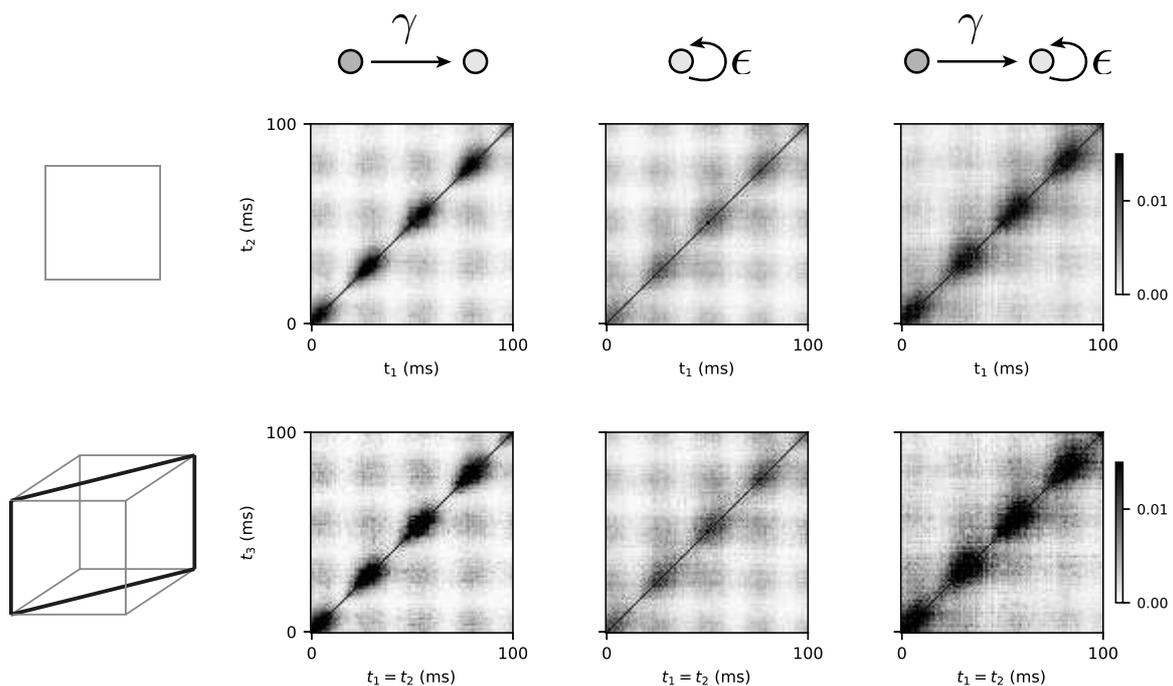}
        	\caption{ \label{fig5_aff_rec}
        	{\bf Comparison between the output moments for various connectivity configurations with a single output neuron.}
	The top matrix row corresponds to the 2nd-order moment and the bottom row to a diagonal plane of the 3rd-order moment.
        The same driving oscillatory intensity is fed to the neuron on the left of each diagram, in particular directly to the output neuron for the middle column.
        We compare the following configurations:
        {\bf Left column:} feedforward network with $w_\mathrm{aff} = 2$ (and $w_\mathrm{rec} = 0$);
        {\bf Middle column:} feedforward network with $w_\mathrm{rec} = 0.5$;
        {\bf Right column:} feedforward network with $w_\mathrm{aff} = 1$ and $w_\mathrm{rec} = 0.5$.
        As before, the plots are results averaged over 10000 simulations.
        }
\end{figure}

\end{example}

Up to now, we have used synaptic kernels with non-negative values, to ensure that the firing intensity in Eq.~\eqref{eq:Hawkes} is always non-negative, which is necessary in our calculations. Now we consider negative synaptic connections to see how our calculations hold despite violating this assumption.
\begin{example}[Two neurons with refractory self-connections and mutual excitation]\label{example:ref}

For the two neurons with self-inhibition in Fig.~\ref{fig6_refrac}A, each output spike triggers a temporary decrease of their firing intensity, thereby implementing relative refractoriness. \red{Fig.~\ref{fig6_refrac}B compares the time courses of the two synaptic kernels for excitation (as used until now) and self-inhibition for refractoriness.}
In order to prevent ``negative firing intensity'' (corresponding to dotted gray curves below the horizontal line at 0 in Fig.~\ref{fig6_refrac}C), we used a rectification function such that the firing intensity remains non-negative, \red{namely $\nu \geq 0$ (solid gray curve). 
This translates to a non-linear non-negative-valued function on the right-hand side of Eq.~\eqref{eq:Hawkes}, here taken as $\nu_i(t) = [\cdots]_+$}.
As mentioned earlier, this situation violates the hypothesis behind our calculations and is expected to result in errors between the predicted moments and their empirical counterparts. As an example, Fig.~\ref{fig6_refrac}D displays the 1st-order moments for the two neurons with $w_\mathrm{refrac} = 0.1$ (as in Fig.~\ref{fig6_refrac}C), which match well the theory. \red{In contrast, the discrepancy with the theory is larger for $w_\mathrm{refrac} = 0.3$ in Fig.~\ref{fig6_refrac}E.}
Note that their 1st-order moments are different for the two neurons, because of the distinct excitatory weights that connect them.
When increasing the refractory weight $w_\mathrm{refrac}$, the firing rate obtained in the simulation deviates from its expected value calculated from the theory: the predicted value is lower because our calculations involve ``negative firing intensity'', as shown for the mean firing rate over the two neurons in Fig.~\ref{fig6_refrac}F. Similarly, the accuracy of the moments decreases for larger $w_\mathrm{refrac}$, corresponding to larger errors in Fig.~\ref{fig6_refrac}G. Note that the error in Fig.~\ref{fig6_refrac}G includes finite size effects, i.e.\ it involves the empirical error due to the simulation over a limited period of time. In general, such errors may also be influenced by the specific choice of the synaptic kernel.

\begin{figure}
        	\includegraphics[scale=1]{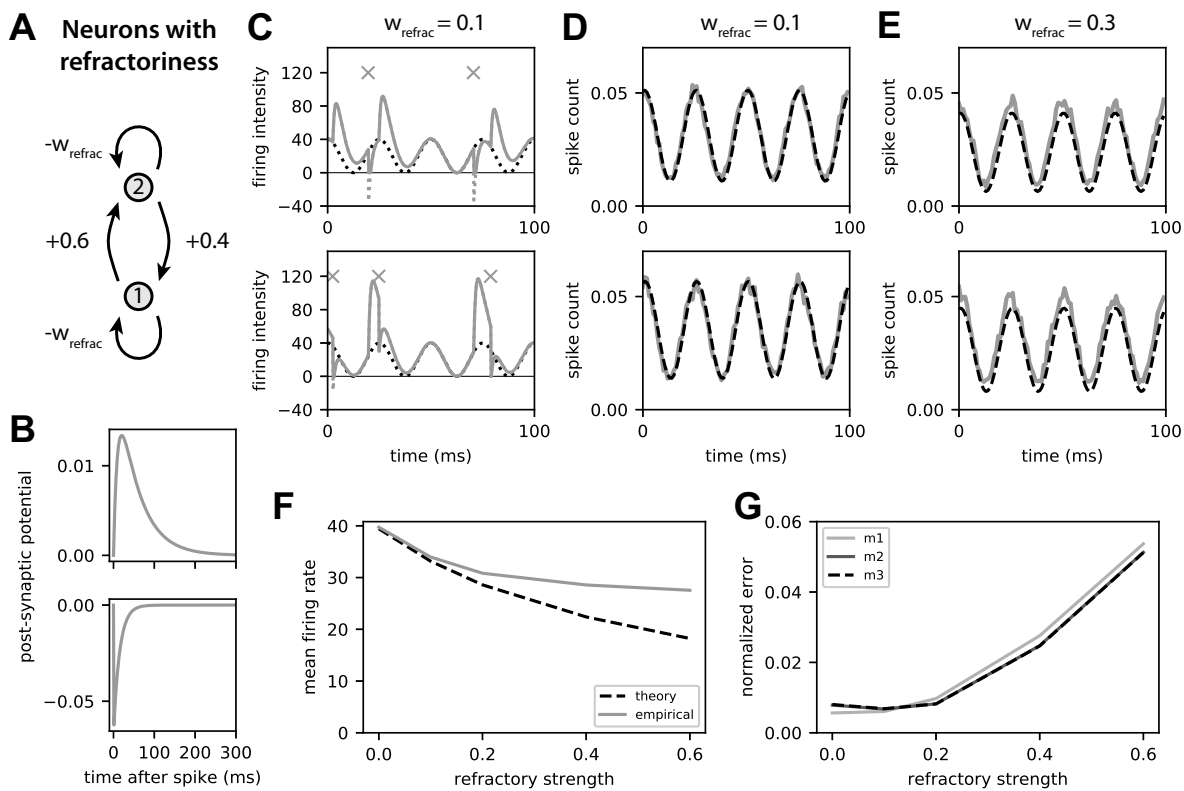}
        	\caption{ \label{fig6_refrac}
        	{\bf Influence of refractoriness on the moment evaluation.}
        {\bf A:} Schematic diagram of two neurons that mutual excite each other with distinct weights and have inhibitory self-connections which model refractoriness.
        {\bf B:} \red{Comparison of the time courses of the kernels used for excitatory and self-inhibitory synapses. The excitatory kernel (top) corresponds to a double exponential $( \exp(-t/\tau_\mathrm{decay}) - \exp(-t/\tau_\mathrm{rise}) ) / (\tau_\mathrm{decay} - \tau_\mathrm{rise})$ with $\tau_\mathrm{rise} = 1$~ms and $\tau_\mathrm{decay} = 5$~ms, as used before in other figures. The inhibitory kernel used for refractoriness (bottom) is a simple decaying exponential $\exp(-t/\tau_\mathrm{refrac}) / \tau_\mathrm{refrac}$ with $\tau_\mathrm{refrac} = 1.5$~ms.}
        {\bf C:} Example firing intensity for the two neurons (1 at bottom and 2 at top) where spikes are indicated by crosses. Here the refractory weight is set to $w_\mathrm{refrac} = 0.1$.
        The driving oscillatory intensity $\lambda$ is represented by the dotted black curve.
        {\bf D:} 1st-order moment (solid gray curve) with theoretical prediction (dashed black curve) for the two neurons and $w_\mathrm{refrac} = 0.1$, averaged over 10000 simulations.
        {\bf E:} \red{Same as panel D for stronger refractoriness with $w_\mathrm{refrac} = 0.3$.}
	{\bf F:} Predicted (dashed black curve) and empirical (solid gray curve) mean firing rate as a function of the refractory weight $w_\mathrm{refrac}$. As before, the results correspond to the average over 10000 repetitions of the same network.
	{\bf G:} Normalized error for the 1st- to 3rd-order moments when varying $w_\mathrm{refrac}$.
	It corresponds to the difference between the theoretical and empirical curves in panel C, squared and integrated over the 100~ms period. The normalization consists in dividing by the firing rate. \red{The non-zero normalized error at  $w_\mathrm{refrac} = 0$ comes from the finite number of simulation repetitions.}
        }
\end{figure}

\end{example}

\section{Relationship with cumulants}

We end with relating our results with previous work~\cite{Jovanovic_PRE_2015,Ocker_PCB_2017} that described the activity in Hawkes networks using cumulants instead of moments.
These studies were limited to the case of neurons driven by deterministic intensities and focused on cumulants because of the theoretical tools that they applied to the present problem, respectively Hawkes branching process and field theory.
Cumulants and moments are two manners to describe the spiking statistics and the genuine relationship between them comes from their generating functions~\cite[Section~5.2]{Balakrishnan_SPL_1998,Daley_1988}.
Let $E^\alpha(\bzeta,\bk,\bt)$ be the moment generating function for the multivariate random variable $\alpha_\bk(\bt) = (\alpha_{k_1}(t_1), \cdots, \alpha_{k_p}(t_p))$:
%
\begin{equation}
    E^x(\bzeta,\bk,\bt) = \Ex{\exp\pa{\sum_{r=1}^{p} \zeta_r \alpha_{k_r}(t_r)}}_\alpha
    \ ,
\end{equation}
where $\bzeta = (\zeta_1, \cdots, \zeta_p)^T$. This moment generating function can be used to express the $p^{\rm th}$ order moment over the coordinates $\bk$ and times $\bt$:
\begin{equation} \label{eq:X_mom_gen}
    X^p_\bk(\bt)
    =
    \frac{\partial^p E^\alpha(\bzeta,\bk,\bt)}{\partial \zeta_{i_1} \cdots \partial \zeta_p} \Big\vert_{\bzeta= 0}
    =
    \frac{\partial^p E^\alpha(\bzeta,\bk,\bt)}{\partial \bzeta} \Big\vert_{\bzeta= 0}
\end{equation}
The cumulant generating function for the random variable $\alpha$ is given by 
\begin{equation} \label{eq:KE}
    K^\alpha(\bzeta,\bk,\bt) = \log E^\alpha(\bzeta,\bk,\bt)
\end{equation}

\begin{notation}[cumulant]
The cumulants of order $p$ for the indices $\bk = (k_1,\dots,k_p)$ at times $\bt = (t_1,\dots,t_p)$ for the input $x$, for the driving intensity $\lambda$ and the filtered inputs $\gamma\ast x$ are defined respectively as
\begin{eqnarray}
\bar{X}^{p}_{\bk}(\bt) &=& \frac{\partial^p K^x(\bzeta,\bk,\bt)}{\partial \bzeta}\Big\vert_{\bzeta= 0}\label{eq:cum_C}\\
\bar{\Lambda}_\bi(\bt) &=& \frac{\partial^p K^\lambda(\bzeta,\bk,\bt)}{\partial \bzeta}\Big\vert_{\bzeta= 0}\label{eq:cum_lambda} \\
\bar{\Gamma}_\bi(\bt) &=& \frac{\partial^p K^{\gamma\ast x}(\bzeta,\bk,\bt)}{\partial \bzeta}\Big\vert_{\bzeta= 0}\label{eq:cum_gamma}\\
\bar{Y}_\bi(\bt) &=&\frac{\partial^p K^y(\bzeta,\bk,\bt)}{\partial \bzeta}\Big\vert_{\bzeta= 0}\label{eq:cum_y}
\end{eqnarray}
\end{notation}
\begin{property}\label{prop:cum_mom}
The formal relationship between the moment $X^p_\bk(\bt)$ of order $p$ and cumulants $\bar{X}^{p'}_{\bk'}(\bt')$ of order $p' \leq p$ ---here presented for the inputs--- is given by
\begin{equation} \label{eq:cum_mom}
    X^p_\bk(\bt)
    =
    \sum_{\Phi \in \mathcal{P}_p} \prod_{S \in \Phi} \bar{X}^{|S|}_{\bk_S}(\bt_S)
    \ ,
\end{equation}
where $\Phi$ are the partitions of $I_p$ composed of disjoint subsets $S$.
\end{property}

\begin{myproof}{Property}{\ref{prop:cum_mom}}
The present proof ---inspired by previous work~\cite{Daley_1988,Balakrishnan_SPL_1998}--- relies on the following general result for the (partial) derivative of $\exp\pa{f}$ with respect to variables $\bzeta = (\zeta_1, \cdots, \zeta_p)$ for an arbitrary function $f$ without specified arguments:
\begin{equation} \label{eq:deriv_part_p}
    \frac{\partial^p \exp\pa{f}}{\partial \bzeta}
    =
    \frac{\partial^p \exp\pa{f}}{\partial \zeta_1 \cdots \partial \zeta_p}
    =
    \pa{ \sum_{\Phi \in \mathcal{P}_p} \prod_{S \in \Phi} \frac{\partial^{|S|} f}{\partial \bzeta_S} } \; \exp\pa{f}
    \ ,
\end{equation}
which involves all partitions $\Phi \in \mathcal{P}_p$ and the partial derivatives $\frac{\partial^{|S|} f}{\partial \bzeta_{S}}$ of order $|S|$ with respect to the variables $\zeta_r$ whose indices $r \in S$.
For $p=1$ with $\zeta_1$, we have the univariate case
\begin{equation} \label{eq:deriv_part_1}
    \frac{\partial \exp\pa{f}}{\partial \zeta_1}
    =
    \frac{\partial f}{\partial \zeta_1} \; \exp\pa{f}
    \ .
\end{equation}
To demonstrate Eq.~\eqref{eq:deriv_part_p}, we assume the expression to be valid for $p-1$ and derive it for $p$, using a proof by induction. Separating $\zeta_p$ from the remaining variables $\bzeta' = (\zeta_1, \cdots, \zeta_{p-1})$, we use Eq.~\eqref{eq:deriv_part_p} for $p-1$:
\begin{eqnarray}
    \frac{\partial^p \exp\pa{f}}{\partial \bzeta}
    & = &
    \frac{\partial}{\partial \zeta_p}\frac{\partial^{p-1} \; \exp\pa{f}}{\partial \bzeta'}
    \nonumber \\
    & = &
    \frac{\partial}{\partial \zeta_p} \pa{ \pa{ \sum_{\Phi \in \mathcal{P}_{p-1}} \prod_{S \in \Phi} \frac{\partial^{|S|} f}{\partial \bzeta_{S}} } \; \exp\pa{f} }
    \nonumber\\
    & = &
    \pa{ \sum_{\Phi \in \mathcal{P}_{p-1}} \frac{\partial}{\partial \zeta_p} \pa{ \prod_{S \in \Phi} \frac{\partial^{|S|} f}{\partial \bzeta_{S}} } } \exp\pa{f}
    \; + \;
    \pa{ \sum_{\Phi \in \mathcal{P}_{p-1}} \prod_{S \in \Phi} \frac{\partial^{|S|} f}{\partial \bzeta_{S}} } \frac{\partial f}{\partial \zeta_p} \exp\pa{f}
    \ ,
\end{eqnarray}
where the derivative with respect to $\zeta_p$ applied to the product yields two terms.
The second term corresponds to Eq.~\eqref{eq:deriv_part_1}, which can be assimilated to the partition $\Phi' \in \mathcal{P}_p$ such that $\Phi' = \Phi \cup \big\{ \pacu{p} \big\}$.
The first term actually gives $|\Phi|$ terms, one for each subset $S$ of the product, which depends on the actual partition $\Phi$. For each $\Phi$, we construct $|S|$ partitions $\Phi' \in \mathcal{P}_p$ by adding the index $p$ to one of the subsets $S \in \Phi$.
Because a partition $\Phi \in \mathcal{P}_p$ can only be of one of the two types, we end up with 
\begin{eqnarray}
    \frac{\partial^p \exp\pa{f}}{\partial \bzeta}
    & = &
    \pa{\sum_{\Phi \in \mathcal{P}_p \setminus \mathcal{Q}_p} \prod_{S \in \Phi} \frac{\partial^{|S|} f}{\partial \bzeta_S}
    +
    \sum_{\Phi \in \mathcal{Q}_p} \prod_{S \in \Phi} \frac{\partial^{|S|} f}{\partial \bzeta_S} } \exp\pa{f}
    \nonumber \\
    & = &
    \pa{ \sum_{\Phi \in \mathcal{P}_p} \prod_{S \in \Phi} \frac{\partial^{|S|} f}{\partial \bzeta_S} } \exp\pa{f}
    \ ,
\end{eqnarray}
where $\mathcal{Q}_p = \pacu{ \Phi \in \mathcal{P}_p, \pacu{p}\in\Phi}$ is the set of all partitions of $I_p$ that contain the singleton $\pacu{p}$.

Coming back to the moments, we prove Eq.~\eqref{eq:cum_mom} by applying Eq.~\eqref{eq:deriv_part_p} to the function $K(\zeta,\bk,\bt)$:
\begin{eqnarray}
    X_{\bk}^p(\bt)
    & = &
    \frac{\partial^p \exp\pa{K(\bzeta,\bk,\bt)}}{\partial \bzeta}\Big\vert_{\bzeta= 0}
    \nonumber \\
    & = &
    \sum_{\Phi \in \mathcal{P}_p}\prod_{S \in \Phi}\frac{\partial^{|S|} K(\bzeta,\bk,\bt)}{\partial \bzeta_S} \exp\pa{K\pa{\bzeta,\bk,\bt}}\Big\vert_{\bzeta= 0}
    \nonumber \\
    & = &
    \sum_{\Phi \in \mathcal{P}_p} \prod_{S \in \Phi} \bar{X}^{|S|}_{\bk_S}(\bt_S)
    \ ,
\end{eqnarray}
after noticing that $\exp\pa{K\pa{{\bf 0},\bk,\bt}} = 1$.
\end{myproof}

\begin{corollary}
A direct corollary of Proposition~\ref{prop:cum_mom} is that, when the input neurons are independent and driven by intensities $\mu_k(t_k)$, then the cumulant of order $p$ of the input population $x$ is given by
\begin{equation} \label{eq:tmp}
    \bar{X}^p_\bk(\bt)
    =
    \deltabar_{\bk}(\bt) \; \mu_{k_1}(t_1)
    \ .
\end{equation}
\end{corollary}
The proof simply consists in identifying the terms in Eq.~\eqref{eq:prop1_input} to the cumulants, where $\bk$ and $\bt$ are respectively replaced by $\bk_S$ and $\bt_S$ for each subset $S$.
\\

Now we examine the general situation of a network with afferent and recurrent connectivities, corresponding to the combined theorems for moments ---see Eqs.~\eqref{eq:mom_rec_uncoupled} and \eqref{eq:net_mapping}.

\begin{theorem} [Mappings for cumulants] \label{th:th_cum}
The cumulants are related by the following mappings:
\begin{subequations}
\begin{eqnarray} \label{eq:mapping_cum1}
    \bar{\Gamma}^{p}_\bi(\bt)
    & = &
    \pa{\bgamma^p_{\bi\bk} \circledast \bar{X}_\bk^p}(\bt) \ ,
    \\ \label{eq:mapping_cum2}
    \bar{Y}^{p,\epsilon=0}_\bi(\bt)
    & = &
    \sum_{\Phi \in \mathcal{P}_p} \pa{ \prod_{S \in \Phi} \deltabar_{\bi_S}(\bt_S) }
    \pa{ \bar{\Gamma}_{\bi_{\check{\Phi}}}^{|\check{\Phi}|}(\bt_{\check{\Phi}}) + \bar{\Lambda}_{\bi_{\check{\Phi}}}^{|\check{\Phi}|}(\bt_{\check{\Phi}}) } \ ,
   \\ \label{eq:mapping_cum3}
    \bar{Y}^{p}_\bi(\bt)
    & = &
    \pa{\btepsilon^p_{\bi\bj} \circledast \bar{Y}^{p,\epsilon=0}_\bj}(\bt)
    \ .
\end{eqnarray}
\end{subequations}
\end{theorem}
\begin{myproof}{Theorem}{\ref{th:th_cum}}
Eq.~\eqref{eq:mapping_cum1} simply comes from the linearity of the filtering by $\gamma$.
Another manner to prove it is to decompose the moment in terms of cumulants, as we do now to demonstrate Eq.~\eqref{eq:mapping_cum3}.\\

By rewriting Eq.~\eqref{eq:net_mapping} in terms of cumulants using Eq.~\eqref{eq:cum_mom}, we have
\begin{eqnarray}
    \sum_{\Phi \in \mathcal{P}_p}\prod_{S \in \Phi} \bar{Y}^{|S|}_{\bi_S}(\bt_S)
    & = &
    \btepsilon_{\bi\bj} \circledast \pa{ \sum_{\Phi \in \mathcal{P}_p} \prod_{S \in \Phi} \bar{Y}_{\bj_S}^{|S|,\epsilon = 0}(\bt_S)}
    \nonumber\\
    & = &
    \sum_{\Phi \in \mathcal{P}_p} \prod_{S \in \Phi} \pa{ \btepsilon_{\bi_S \bj_S} \circledast \bar{Y}_{\bj_S}^{|S|,\epsilon=0}}(\bt_S)
    \ .
\end{eqnarray}
As before, we identify the terms for each $S$ and $\Phi$.\\

In contrast, Eq.~\eqref{eq:mapping_cum2} is not straightforward and comes from the spiking nature of $y$ driven by an intensity function $\nu^{\epsilon = 0}$ that possibly has high-order correlations (for example a Cox process).
Basically, it is the extension of cumulants of smaller orders by delta functions for all possible partitions for each time variable of the smaller-order cumulant.
For simplicity, we only show the result for $\bar{\Gamma}$; note also that the additivity of the cumulant ensures the complete result. 
We rewrite Eq.~\eqref{eq:mom_no_lambda} ---that is the equivalent of Eq.~\eqref{eq:mom_rec_uncoupled} in the absence of $\lambda$--- in terms of cumulants using Eq.~\eqref{eq:cum_mom}:
\begin{equation} \label{eq:dev_cum1}
    \sum_{\Phi \in \mathcal{P}_p} \prod_{S \in \Phi} \bar{Y}^{|S|,\epsilon=0}_{\bi_S}(\bt_S)
    =
    \sum_{\Phi \in \mathcal{P}_p}
    \pa{\prod_{S \in \Phi} \deltabar_{\bi_S}(\bt_S)} \pa{\sum_{\Phi' \in \mathcal{P}(\check{\Phi})} \prod_{S' \in \Phi'} \bar{\Gamma}^{|S'|}_{\bi_{S'}}(\bt_{S'})}
    \ .
\end{equation}
In Eq.~\eqref{eq:dev_cum1} cumulants $\bar{\Gamma}$ involve indices from distinct subsets $S$ of the partition $\Phi$, as they ``combine'' the minima in $\check{\Phi}$ according to $\Phi'$.
We now reorganize the expression to obtain a similar expression to the left-hand side, where the terms in the product over $S$ have a generic expression with indices only in $S$. 
The product of generalized delta functions can be moved inside the sum over $\Phi'$, yielding
\begin{equation} \label{eq:dev_cum2}
    \sum_{\Phi \in \mathcal{P}_p} \prod_{S \in \Phi} \bar{Y}^{|S|,\epsilon=0}_{\bi_S}(\bt_S)
    =
    \sum_{\Phi \in \mathcal{P}_p} \sum_{\Phi' \in \mathcal{P}(\check{\Phi})}
    \pa{\prod_{S \in \Phi} \deltabar_{\bi_S}(\bt_S)}
    \pa{\prod_{S' \in \Phi'} \bar{\Gamma}^{|S'|}_{\bi_{S'}}(\bt_{S'})}
    \ .
\end{equation}
For each pair of partitions $\Phi$ and $\Phi'$, we construct a partition $\Psi \in \mathcal{P}_p$, whose subsets $T$ are the unions of subsets $S$ corresponding to the same $S' \in \Phi'$:
\begin{equation} \label{eq:def_T}
    T
    =
    \bigcup_{\substack{S \in \Phi, \\ \check{S} \in S' \in \Phi'}} S
    \ .
\end{equation}
In addition, we define a partition $\Psi'_T \in \mathcal{P}(T)$ for each $T \in \Psi$ that splits $T$ into the original subsets $S \in \Phi$:
\begin{equation} \label{eq:def_psiprime}
    \Psi'_T
    =
    \bigcup_{S \in T} \pacu{ S } 
    \ .
\end{equation}

The correspondence between the partitions is represented in Fig.~\ref{fig7_partitions} for a schematic example.
Using Eq.~\eqref{eq:def_psiprime} with $S = T' \in \Psi'_T$ for the each $T$, the first product in Eq.~\eqref{eq:dev_cum2} can be rewritten as
\begin{equation} \label{eq:dev_cum3}
    \prod_{S \in \Phi} \deltabar_{\bi_S}(\bt_S)
    =
    \prod_{T \in \Psi} \; \prod_{T' \in \Psi'_T} \deltabar_{\bi_{T'}}(\bt_{T'})
    \ .
\end{equation}
Because each $S' \in \Phi' = \check{\Phi}$ is the subset of minima $\check{\Psi}'_T$ for the corresponding $T = \bigcup S$, we similarly reformulate the second product
\begin{equation} \label{eq:dev_cum4}
    \prod_{S' \in \Phi'} \bar{\Gamma}^{|S'|}_{\bi_{S'}}(\bt_{S'})
    =
    \prod_{T \in \Psi} \bar{\Gamma}^{|\check{\Psi}'_T|}_{\bi_{\check{\Psi}'_T}}(\bt_{\check{\Psi}'_T})
    \ .
\end{equation}
We can thus factorize the two products in the right-hand side of Eq.~\eqref{eq:dev_cum2} to obtain
\begin{equation} \label{eq:dev_cum5}
    \pa{\prod_{S \in \Phi} \deltabar_{\bi_S}(\bt_S)}
    \pa{\prod_{S' \in \Phi'} \bar{\Gamma}^{|S'|}_{\bi_{S'}}(\bt_{S'})}
    =
    \prod_{T \in \Psi} \; \pa{ \prod_{T' \in \Psi'_T} \deltabar_{\bi_{\check{T}'}}(\bt_{\check{T}'}) }
    \bar{\Gamma}^{|\check{\Psi}'_T|}_{\bi_{\check{\Psi}'_T}}(\bt_{\check{\Psi}'_T})
    \ .
\end{equation}
Last, the key observation is that each pair $\Phi \in \mathcal{P}_p$ and $\Phi' \in \mathcal{P}(\check{\Phi})$ is uniquely associated with another pair made of a partition $\Psi \in \mathcal{P}_p$ and its corresponding set of partitions $\pacu{\Psi'_T\in\mathcal{P}(T)}_{T\in\Psi}$:
\begin{equation}
    (\Phi,\Phi') \leftrightarrow (\Psi,\pacu{\Psi'_T}_{T\in\Psi})
    \ .
\end{equation}
As a consequence, the double summation over $\Phi$ and $\Phi'$ in Eq.~\eqref{eq:dev_cum2} can be expressed as a summation over  $\Psi$ and over its corresponding sub-partitions, namely
\begin{equation} \label{eq:swap_sum}
    \sum_{\Phi \in \mathcal{P}_p} \sum_{\Phi' \in \mathcal{P}(\check{\Phi})}
    \leftrightarrow
    \sum_{\Psi \in \mathcal{P}_p} \sum_{\Psi'_{T_1} \in \mathcal{P}(T_1)} \cdots \sum_{\Psi'_{T_{|\Psi|}} \in \mathcal{P}(T_{|\Psi|})}
    = 
    \sum_{\Psi \in \mathcal{P}_p} \; \sum_{\substack{\Psi'_T \in \mathcal{P}(T), \\ \forall T \in \Psi}}
    \ ,
\end{equation}
with the explicit enumeration of $T_r \in \Psi$.
With this substitution, Eq.~\eqref{eq:dev_cum2} can be expressed as:
\begin{eqnarray} \label{eq:dev_cum6}
    \sum_{\Phi \in \mathcal{P}_p} \prod_{S \in \Phi} \bar{Y}^{|S|,\epsilon=0}_{\bi_S}(\bt_S)
    & = &
    \sum_{\Psi \in \mathcal{P}_p} \; \sum_{\substack{\Psi'_T \in \mathcal{P}(T), \\ \forall T \in \Psi}} \; \prod_{T \in \Psi}
    \pa{ \prod_{T' \in \Psi'_T} \deltabar_{\bi_{T'}}(\bt_{T'})} \bar{\Gamma}^{|\check{\Psi}'_T|}_{\bi_{\check{\Psi}'_T}}(\bt_{\check{\Psi}'_T})
    \nonumber \\
    & = &
    \sum_{\Psi \in \mathcal{P}_p} \prod_{T \in \Psi} \pa{ \sum_{\Psi' \in \mathcal{P}(T)}
    \pa{ \prod_{T' \in \Psi'} \deltabar_{\bi_{T'}}(\bt_{T'})} \bar{\Gamma}^{|\check{\Psi}'|}_{\bi_{\check{\Psi}'}}(\bt_{\check{\Psi}'}) }
    \ .
\end{eqnarray}
Once again, we conclude by identifying the terms for each $T$ and $\Psi$ in the right-hand side and $S$ and $\Phi$ in the left-hand side of Eq.~\eqref{eq:dev_cum6}.

\begin{figure}
        	\includegraphics[scale=1]{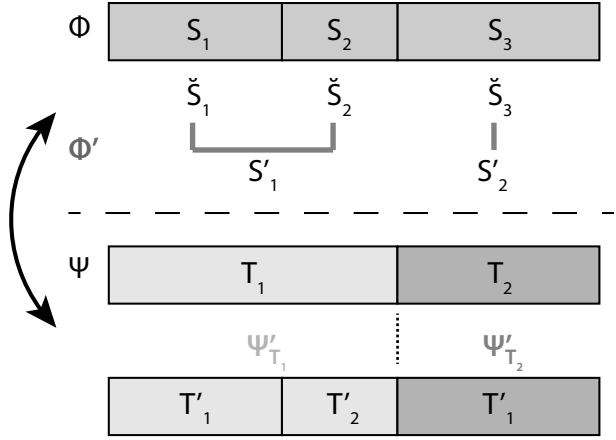}
        	\caption{ \label{fig7_partitions}
        	{\bf One-to-one mapping between partitions.}
	The partition $\Psi$ is constructed from the pair $\Phi$ and $\Phi'$. To each element of $T \in \Psi$ corresponds a partition $\Psi'_{T}$ that recovers the original subsets in $\Phi$. See Eqs.~\eqref{eq:def_T} and \eqref{eq:def_psiprime} in the main text for the mathematical construction. Here the subsets are indexed as in Eq.~\eqref{eq:swap_sum} and the partitions $\Psi'_{T}$ are represented using different gray contrasts.
        	}
\end{figure}
\end{myproof}

Note that Eq.~\eqref{eq:mapping_cum2} for cumulants resembles its counterpart Eq.~\eqref{eq:mom_rec_uncoupled} for moments, but in the case where the neurons are stimulated by both inputs and driving intensities, the corresponding cumulants are simply summed, whereas moments appear in a product.

\section{Discussion}

In this paper we analytically computed the statistics of neuronal activity in a recurrent network ---described via moments and then transposed to cumulants--- from the statistics of the input neuronal population. An important contribution of our study is the description of the propagation of spiking moments in feedforward networks (Theorem~\ref{th:th_aff}) and recurrently-connected networks (Theorem~\ref{th:th_rec}), which had not been explored before. Theorem~\ref{th:th_cum} established the equivalent mappings for cumulants. Compared to recent studies for cumulants~\cite{Jovanovic_PRE_2015,Ocker_PCB_2017}, an important advantage of the operator viewpoint taken here is that it provides intuition about the spatio-temporal filtering induced by both afferent and recurrent connectivities. In particular, Fig.~\ref{fig5_aff_rec} shows that the combination of afferent and recurrent connectivities can lead to strong output correlation structure, hinting at nonlinear effects on the distribution of spikes. This can be explained by the recurrent connectivity, as well as the interplay between moment orders (here from low-order to high-order moments).
Another interesting point is that moments do not require a stationarity assumption to derive their consistency equation, in contrast to covariances in the original Hawkes' formalism.

The main technical challenge comes from the spiking nature of neurons which forces us to consider all possible contractions, see Eq.~\eqref{eq:dNp}. For rate-based neurons ---still interacting through spatio-temporal kernels--- or equivalently assuming that the population size is very large such that individual spikes have negligible effects, our results can be expressed in a much simpler way (see Remark~\ref{rem:approx}). In this case, the output moments can be approximated by a nested convolution: a first convolution of the input moments with the feedforward kernel followed by a second convolution with the effective recurrent kernel. Quantifying the deviations from this approximation for neuronal population of finite size is left for future work.

At the heart of the tractability in this study is the linearity assumption of the Hawkes process considered here, as the firing probability is proportional to the membrane potential that simply sums the synaptic inputs. This allows us to calculate exact results that do not rely on, e.g., the diffusion approximation.
However, the linearity assumption obviously imposes limitations to the scope of the theory presented here. In particular, refractoriness (i.e.\ the reduction of spiking probability during the few milliseconds that follow an action potential) as well as inhibition, which is ubiquitous in the brain, cannot be exactly modeled in this linear framework. 
\red{Discrepancies with the theory will of course depend on many factors such as the shapes and amplitudes of the synaptic kernels (especially for inhibitory connections) as well as the properties of the filtered inputs and driving intensities ---which may interplay.} Nonetheless, we have found that our formalism still holds when such mechanisms modeled by negative connectivity weights are not too strong (Fig.~\ref{fig6_refrac}).

To circumvent those limitations, several studies included various forms of nonlinearities in the Hawkes process ~\cite{Bremaud_AP_1996,Galves_JSFS_2016,Chevallier_SPA_2017,Gao_SPA_2018,Ferrari_JSP_2018,Raad_arXiv_2018}. One can assume that the firing intensity explicitly depends on the time difference up to the previous spike in order to make the Hawkes process age-dependent~\cite{Raad_arXiv_2018}. This approach ensures that the stability is independent of the two classical stability conditions, i.e.\ (a) $\alpha$-Lipschitz condition on the nonlinear intensity function and (b) the integral of the absolute value of the recurrent kernel should be smaller than $1/\alpha$~\cite{Bremaud_AP_1996}. However, the computation of moments and cumulants in this context is expected to be much harder. Mean-field approximations lead to analytical results~\cite{Toyoizumi_NC_2009}, but they are only valid in the limit of weak coupling. 
Another possibility is to rely on path-integral formulation and the related Feynman diagram formalism~\cite{Shchepanyuk_UKJ_1995,Ocker_PCB_2017,Chen_arxiv}, but it brings an additional complexity that requires further analysis to obtain an intuitive understanding of the combined effect of the feedforward and recurrent kernels in propagating spiking moments. 
In that respect, the linear Hawkes process already leads to complex cross-overs between cumulants ---as can be seen in Eq.~\eqref{eq:mapping_cum2}--- and many more are expected to appear for the nonlinear case. Note that gaining insight about this cross-talk between moment orders is important to investigate a network driven by correlated inputs or by driving intensities with correlation structure such as Cox processes~\cite{Lechnerova_MCAP_2008,Laier_AAP_2008}.

In the context of neuroscience, our results can be applied to the field of synaptic plasticity under two conditions. First, the output neurons should be in the linear regime and secondly the learning rule should have a small learning rate (i.e.\ learning operates at a much slower time scale than neuronal dynamics). For activity-dependent models, the expected weight change can be expressed from the corresponding statistics of the spiking activity~\cite{Kempter_PRE_1999,Gilson_FCN_2010}. Furthermore, since synaptic plasticity has been demonstrated to depend on higher-order correlations~\cite{Pfister_JNS_2006,Clopath_NN_2010}, our formalism provides the adequate tools to analytically study synaptic plasticity in recurrently-connected networks, extending previous work that relied on approximations~\cite{Gjorgjieva_PNAS_2011}.

Efforts have been made to fit univariate Hawkes processes to empirical time series using Bayesian estimation based on the likelihood~\cite{Ozaki_AISM_1979,Truccolo_JPP_2016,Laub_arxiv,Fujita_arxiv}, relative spike count between neurons~\cite{Lambert_JNM_2018} or relying on average second-order statistics~\cite{Fonseca_JFM_2014,Bacry_IEEE_2016}. Refinements have also been explored in the case of sparse observations of the network activity over time~\cite{Le_IEEE_2018}.
It remains to be explored whether high-order moments can be useful for parameter estimation.

Last, the difference between the abstract space and time in the spiking activity ---namely the coordinates of $x_k(t)$ and $y_i(t)$--- is simply their discrete and continuous natures. The moments tensors could also be defined with continuous space-time variables, adapting Eq.~\eqref{eq:def_tensor_convol} with a spatial integral in line with previous work~\cite{Moller_SPL_2007}. Because our proof relies on linear algebra, it can easily be extended to this new context. The equivalence of the roles of space and time can be seen in the `Generalized spatio-temporal delta function' in Definition~\ref{def:gen_delta} and in the `Matrix convolution' in Definition~\ref{def:mat_convol}.
Analogies with other processes have also been made, such as with the integer-value autoregressive process~\cite{Kirchner_SPA_2016}. It remains to be explored whether such formal mappings between processes provide intuition to interpret these dynamic systems.


\begin{appendix}

\section{Appendix} \label{an:op_aff}

\begin{definition}[Moment symmetrical expansion operator]
Let us consider two tensors of order $q$ and $r$, say $T^q_{\bj'}(\bt')$ with coordinates $\bj' = (j'_1, \cdots j'_{q})$ and $\bt' = (t'_1, \cdots t'_{q})$ as well as $U^r_{\bj''}(\bt'')$ with coordinates $\bj'' = (j''_1, \cdots j''_{q})$ and $\bt'' = (t''_1, \cdots t''_{r})$.
For any given $p \geq q+r$, we define the following tensor operation that constructs a moment of order $p$ with $\bi = (i_1, \cdots, i_p)$ and $\bt = (t_1, \cdots, t_p)$ from the tensors $T$ and $U$ of smaller orders $q$ and $r$:
\begin{equation} \label{eq:tensor_augm}
    \mathcal{A}^p[T^q,U^r]_\bi(\bt)
    =
    \sum_{\substack{ A \subset I_p, B \subset I_p \\ |A| = q, |B| = r\\ A \cap B = \emptyset }} \ \sum_{\substack{\Phi \in \mathcal{P}_p \\ \check{\Phi} = A \cup B}} \pa{\prod_{S \in \Phi} \deltabar_{\bi_S}(\bt_S) } \; T^q_{\bi_A}(\bt_A) \; U^r_{\bi_B}(\bt_B)
    \ .
\end{equation}
Recall that $\check{\Phi} = \{ \check{S}, S \in \Phi \}$ is the set of minima for the groups in the partition $\Phi$.
By convention, the 0-order tensors are valued 1 when $A$ or $B = \emptyset$. 
\end{definition}
Eq.~\eqref{eq:tensor_augm} uses contractions to augment the order of the combinations of tensors $T^q$ and $U^r$ from $q+r$ to $p$ with all possible symmetries. In particular, if $T^q$ and $U^r$ are symmetric tensors (see Remark~\ref{rem:sym_mom}) with respect to all their own dimensions, the output of $\mathcal{A}^p$ is symmetric as well.

Here we reformulate the result of Theorem~\ref{th:th_aff} to group moments of the same order together, using the operator defined in Eq.~\eqref{eq:tensor_augm}.
From Eq.~\eqref{eq:res_th1}, we swap the summation terms of the partitions $\Phi$ and the decomposition of $\check{\Phi}$ in two subsets.
\begin{equation}
      Y^{p,\epsilon=0}_\bi(\bt)
      =
      \sum_{\substack{ A \subset I_p, B \subset I_p \\ A \cap B = \emptyset }} \ \sum_{\substack{\Phi \in \mathcal{P}_p\\ \check{\Phi} = A \cup B}}
      \pa{\prod_{S \in \Phi} \deltabar_{\bi_S}(\bt_S)} \Gamma^{|A|}_{\bi_A}(\bt_A) \ \Lambda^{|B|}_{\bi_B}(\bt_B)
      \ .
\end{equation}
The important point here is to understand that the construction of $\check{\Phi}$ from $A$ and $B$ exactly spans the whole set of partitions $\mathcal{P}_p$.
Note also that $A$ and $B$ can be empty sets.
Then we simply group the subsets $A$ of the same size $q$, and similarly $B$ of the same size $r$:
\begin{equation}
      Y^{p,\epsilon=0}_\bi(\bt)
      =
      \sum_{0 \leq q + r \leq p} \sum_{\substack{ A \subset I_p, B \subset I_p\\ |A| = q, |B| = r \\ A \cap B = \emptyset }} \ \sum_{\substack{\Phi \in \mathcal{P}_p\\ \check{\Phi} = A \cup B}}
      \pa{\prod_{S \in \Phi} \deltabar_{\bi_S}(\bt_S)} \Gamma^{q}_{\bi_A}(\bt_A) \ \Lambda^{r}_{\bi_B}(\bt_B)
    \ ,
\end{equation}
which gives a reformulation of Eq.~\eqref{eq:mom_rec_uncoupled} using the operator $\mathcal{A}^p$ in Eq.~\eqref{eq:tensor_augm}.

\end{appendix}

\section*{Acknowledgments}

MG acknowledges funding from European Union's Horizon 2020 research and innovation programme via the Marie Sk{\l}odowska-Curie Action (H2020-MSCA-656547) and under Grant Agreement No.~785907 (HBP SGA2).
JPP was supported by the Swiss National Science Foundation (SNSF) grants PP00P3\_150637 and PP00P3\_179060.


\end{document}